\documentclass{article}
\usepackage{jcappub}
\usepackage{graphicx} 
\usepackage{dirtytalk}

\usepackage{hyperref}
\usepackage{amssymb}
\usepackage{subcaption}

\usepackage[font=Large,labelfont=bf]{caption}
\usepackage{amsmath,xcolor}
\usepackage[nameinlink, capitalise]{cleveref}
\usepackage{orcidlink}

\usepackage{letltxmacro}
\LetLtxMacro{\originaleqref}{\eqref}
\usepackage{appendix}
\usepackage[sorting=none]{biblatex} 
\RequirePackage[top=20mm,bottom=20mm,left=30mm,right=30mm,head=15mm,includeheadfoot]{geometry}

\title{Constraining the impact of standard model phase transitions on primordial black holes}

\author{Xavier Pritchard\,\orcidlink{0009-0007-6543-8563} and}
\author{Christian T.~Byrnes\,\orcidlink{0000-0003-2583-6536}}

\date{October 2023}

\affiliation{Department of Physics and Astronomy, University of Sussex, Brighton BN1 9QH, UK\\}

\emailAdd{X.Pritchard@sussex.ac.uk}
\emailAdd{C.Byrnes@sussex.ac.uk}

\abstract{Phase transitions in the early universe lead to a reduction in the equation of state of the primordial plasma. This exponentially enhances the formation rate of primordial black holes. However, this sensitivity to the equation of state is the same that primordial black hole abundances show to the primordial curvature power spectrum amplitude. In this paper, we investigate peaked power spectra and show the challenges associated with motivating populations of primordial black holes with standard model enhancements. The parametrisation of different power spectra plays an important role in this discussion. The allowed parameter space consistent with a large QCD phase transition impact on the primordial black hole abundance differs greatly. This is particularly evident for broader spectra. We also show that, in our framework, the electroweak phase transition cannot significantly affect the overall abundance.}
\begin{document}

\maketitle

\section{Introduction}

The idea that compact objects in the early universe could have observational consequences in the present day was first postulated over fifty years ago \cite{Zeldovich}. Shortly after this work, it was suggested that these objects could be black holes, formed from the collapse of density perturbations \cite{Hawking,Carr1}. And so was born the field of primordial black holes (PBHs).

Relatively early on in the field's lifetime, it was known that PBHs were a possible dark matter candidate \cite{Chapline} (see \cite{Green,Carr2} for reviews). More recently, there have been several advancements in gravitational wave (GW) science, such as the Laser Interferometer Gravitational-Wave Observatory (LIGO), Virgo and the Kamioka Gravitational Wave Detector (KAGRA) \cite{Abbott,Abbott1,KAGRA:2013rdx}. These advancements have led to new perspectives on PBHs, such as accounting for the detected GWs from BH mergers \cite{Bird,Franciolini,Escriva1}. Amongst several other factors, this has led to a huge surge in the literature (see \cite{bagui,Escriva2,yoo} for recent reviews of the field).

PBH studies typically require an understanding of primordial curvature perturbations. This information is encoded in the primordial curvature power spectrum $\mathcal{P}_\zeta(k)$, assuming the perturbations follow a Gaussian distribution (for non-Gaussian PBH studies see \cite{Atal:2018neu,Gow1,pi}, for example). In the context of PBHs, $\mathcal{P}$$_\zeta$ needs to be enhanced by many orders of magnitude from the amplitude measured on the large scales ($10^{-4}$ Mpc$^{-1}$ $\lesssim$ $k$ $\lesssim$ 1 Mpc$^{-1}$) of the cosmic microwave background (CMB), where the power spectrum has been measured with extreme accuracy \cite{32}.

A popular model used to achieve such an enhancement, whilst matching CMB observations, is ultra-slow-roll inflation \cite{35,36,Heydari:2023xts}. The level of fine-tuning needed to achieve such a boost of power on small scales is significant in several theoretically motivated single-field inflation models \cite{37} (see also \cite{Qin,Stamou} for more promising multi-field studies). Regardless, we assume there is a peak and study how different parametrisations (and thus shapes) of the peak impact PBH abundances and constraints.

PBH studies not only require an understanding of curvature perturbations in the early universe but also of the universe itself when these perturbations re-enter the horizon. Typically, PBHs are thought to form when the universe is radiation dominated (see \cite{Harada1,Harada2,De_Luca} for examples of PBHs forming during an early matter dominated epoch and \cite{Allahverdi:2020bys} for a review). The pressure of said radiation is dictated by the standard model (SM) of particle physics. As the universe expands and cools, particles become nonrelativistic, and the pressure of the radiation drops more rapidly than the energy density. 

It has been understood for a long time that the reduction in pressure of the primordial plasma leads to an exponential enhancement in the formation rate of PBHs \cite{Chapline,PhysRevD.55.R5871,Cardall:1998ne,Jedamzik:1999am}. Making use of this SM-related enhancement and assuming a flat $\mathcal{P}_\zeta(k)$, it was argued that several cosmological questions such as LIGO-Virgo-KAGRA observations and the origin of supermassive black holes could be explained in a unified PBH scenario \cite{cosmicconundra}. This scenario has since been shown to be unviable \cite{Serpico}, or at the very least it is finely-tuned. One of the main goals of this paper is to extend this investigation to peaked power spectrum. Thus, we will consider whether the SM-related enhancement in abundance can motivate answers to the originally considered cosmic conundra for more realistic power spectra.

A difficulty often faced when motivating PBHs at a given mass scale is the sensitivity of PBH abundance to the amplitude of $\mathcal{P}$$_\zeta$, which is also exponential. Hence, we expect PBH formation to take place only on scales very close to where the primordial power spectrum peaks, regardless of whether this coincides with a scale at which a SM particle decouples. If these scales don't coincide but you would like a population of PBHs motivated by the SM, you must then have a sufficiently broad power spectrum. Thus, we investigate the challenges associated with having such a broad peak.

The key challenge to generating PBHs on larger scales than the horizon scale during the QCD transition are CMB spectral distortions, whose constraints on $\mathcal{P}$$_\zeta$ are about three orders of magnitude tighter than observational PBH constraints \cite{Chluba:2012we}. Thus, power spectra peaked at these scales with broad tails to small $k$ are ruled out. On smaller scales, pulsar timing array (PTA) constraints from the possible induced stochastic GWs are comparably tight to direct PBH constraints. This places a less tight constraint on the tails of the primordial power spectrum towards smaller scales. 

We find that power spectrum constraints are largely dependent on the parametrisation you assume. For some parametrisations of $\mathcal{P}$$_\zeta$, motivating a population of PBHs at a given SM-related scale through the associated reduction in pressure is tightly constrained. This requires the power spectrum to be peaked at this scale, which we argue is fine-tuned. For other shapes of the power spectrum we find that the standard model is able to have a large impact on primordial black hole abundances. 

The paper is organised as follows. In \cref{section 2} we review early universe phase transitions, tracking both the pressure and energy density of the primordial plasma and thus the equation of state and threshold for collapse. In \cref{section 3} we describe the power spectrum parametrisations that we consider, before explaining how we calculate our PBH abundances. In \cref{section 4} we summarise the current constraints on the primordial power spectrum which we then use to constrain the parameters of our chosen power spectra. Throughout this section, we pay special attention to the mass scales relevant to the SM. We conclude our findings in \cref{section 5}. Finally, we discuss additional details and extensions of our work in the appendix.

\section{Relativistic degrees of freedom in the early universe}\label{section 2}

Reheating leaves the universe filled with a hot, dense plasma of relativistic SM particles \cite{Abbott2,Traschen} (see \cite{Bassett,Allahverdi,Jedamzik:2024wtq} for reviews). As the universe expands the density and pressure of the plasma decrease at the same rate, along with the temperature. As the temperature decreases below the mass threshold of a given SM particle the particle decouples from the plasma, resulting in a drop in relativistic degrees of freedom. This results in the ratio between pressure $p$ and the total energy density $\rho$, or equation of state, becoming temporarily non-constant and varying with temperature as 
\begin{equation}\label{EoS}
\omega(T)\equiv\frac{p(T)}{\rho(T)}.
\end{equation}
To obtain our early universe equation of state we use the following formulas for $\rho_i$ and $p_i$ for a given particle, $i$, in thermal equilibrium 
\begin{align}\label{lowtempEoS}
    \rho_i&=\frac{g_i}{2\pi^2}\int_{m_i}^\infty \frac{\sqrt{E^2-m_i^2}}{\text{exp}(E/T)\pm1}E^2\text{d}E,\\
     &p_i=\frac{g_i}{6\pi^2}\int_{m_i}^\infty\frac{(E^2-m_i^2)^{3/2}}{\text{exp}(E/T)\pm1}\text{d}E,
\end{align}
where $m_i$ is the particle’s mass, $g_i$ is the particle's degrees of freedom, and the plus (minus) sign in the denominator applies to fermions (bosons). This information can be found in \cite{Lars}, for instance. To obtain the total pressure and density and hence $\omega(T)$ we sum over all particles.

To relate temperature to horizon mass, wavenumber and time we use the following approximate relations \cite{Nakama1,Carr:2009jm}
\begin{equation}\label{horizonmass}
\begin{split}
 M_H& \simeq 1.5\times10^5\left(\frac{g_{\ast\rho}}{10.75}\right)^{-1/2}\left(\frac{T}{\text{1MeV}}\right)^{-2}M_\odot  ,\\ 
 & \simeq  17\left(\frac{g_{\ast\rho}}{10.75}\right)^{-1/6}\left(\frac{k}{10^6\text{Mpc}^{-1}}\right)^{-2}M_\odot,\\
 & \simeq 2\times10^5\left(\frac{t}{1\text{s}}\right)M_\odot,
\end{split}
\end{equation}
where $g_{\ast\rho}$ is the number of relativistic degrees of energy density as a function of $T$ and $k$ respectively. We note the cancellation of $g_{\ast\rho}$ contributions in the final equation.  

\subsection{Standard model phase transitions}\label{SM PTs}

\begin{figure}[!ht]
    \centering
    \includegraphics[width=\linewidth]{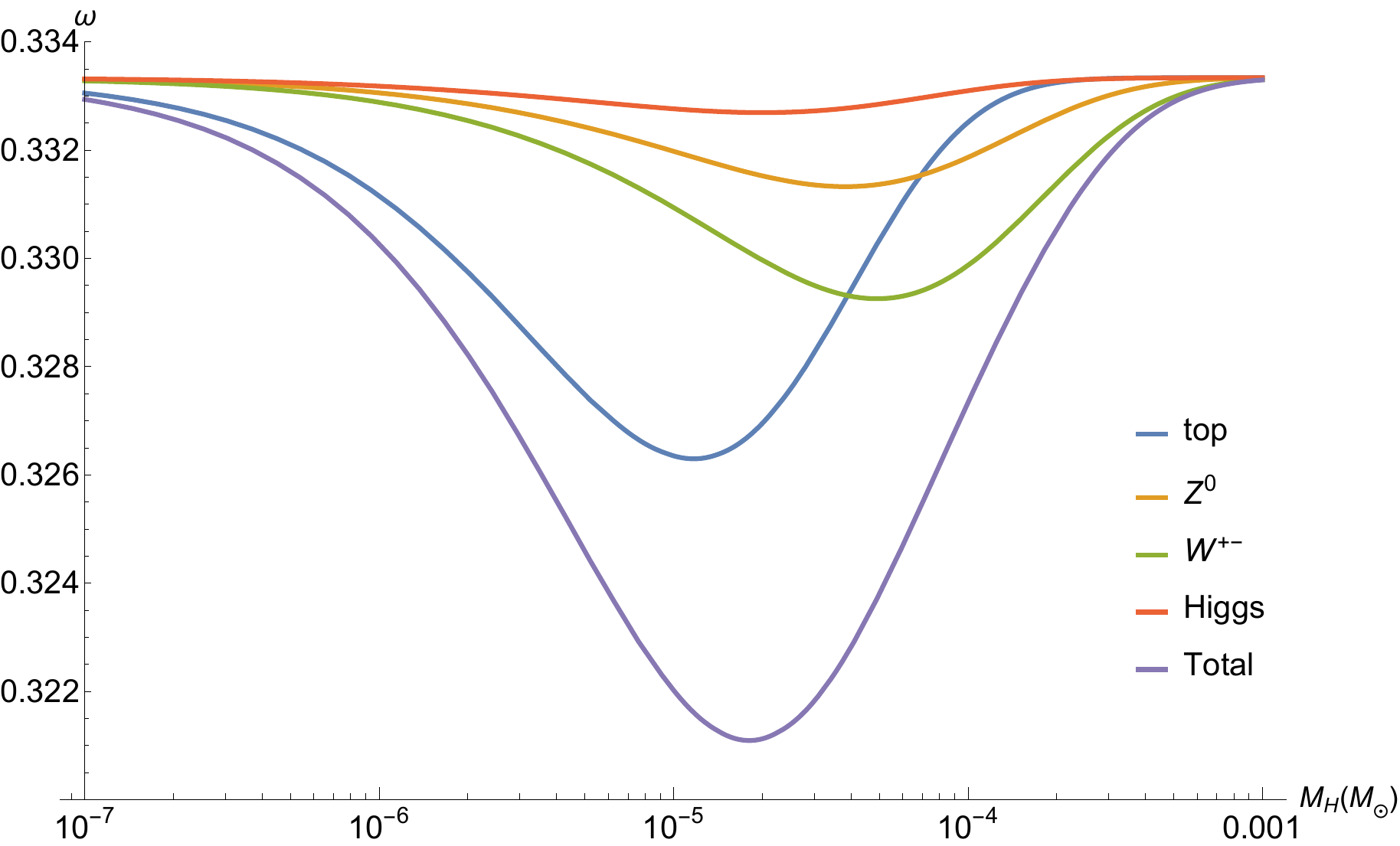}
    \caption{Total reduction in the equation of state $\omega$ during the EW crossover, as a function of the horizon mass of the universe. We also track each relevant particle's contribution to this reduction.}
    \label{EWparticles}
\end{figure}

The first drop in the equation of state in relation to SM particles occurs at around 100 GeV (corresponding to $M_H\approx10^{-5}$ $M_\odot$) when the electroweak (EW) gauge symmetry SU(2)$\times$U(1) is spontaneously broken by the vacuum expectation value of the Higgs field. The observed mass of the Higgs boson \cite{Aad} suggests that this symmetry breaking is a smooth crossover rather than a first-order phase transition \cite{Kajantie}, thus this period is named the electroweak crossover. During this crossover the top quark, Z, W, and Higgs bosons each become non-relativistic. In \cref{EWparticles} we track the contributions to $\omega$ during the EW PT as an example. 

In minimal extensions of the standard model, the EW PT can be strongly first order, for example in Next-to-Minimal supersymmetry models \cite{Huber,Huber2,Demidov}, composite Higgs models \cite{bruggisser,Bian} and many more. This would result in violent bubble collisions and the scalar order parameter and fluid of light particles could produce GWs \cite{Hindmarsh} within the sensitivity region of future GW detectors.

The most drastic drop in $\omega(T)$ occurs during the QCD PT when the temperature of the universe is approximately 100 MeV(corresponding to $M_H\approx1$ $M_\odot$). This occurs when strong interactions confine quarks into hadrons. Lattice QCD studies suggest that similar to the EW PT, the QCD PT is also a crossover \cite{Bazavov,Borsanyi}. Lattice studies are needed due to QCD being highly non-perturbative. To provide a complete equation of state, we thus interpolate the results of \cite{borsanyi1}.

Shortly after the QCD PT pions and muons decouple from the plasma. Finally, at around 1 MeV (corresponding to $M_H\approx10^6$ $M_\odot$), $e^-e^+$ annihilation and neutrino decoupling occur. It has recently been argued that, in the context of PBH formation, the reduction in $\omega(T)$ during the $e^-e^+$ annihilation epoch doesn't necessarily enhance abundances \cite{musco,Jedamzik:2024wtq}. The argument is that, since neutrino interactions with the rest of the plasma freeze out shortly before this PT, they are able to free-stream out of overdense regions. 

\subsection{Threshold for collapse}

The critical threshold for collapse, or $\delta_c$ as it is often written, is a key quantity when calculating PBH abundances. This quantity defines the minimum value a density perturbation amplitude must have at horizon entry for a BH horizon to form and collapse. Generally, a black hole collapses if the inwards gravitational force overcomes the outwards force of pressure gradients. In our context, the density (pressure) of the primordial plasma when a perturbation re-enters the horizon will set the gravitational (pressure) forces. Thus, it is naturally a function of the equation of state $\omega$. We note that the collapse threshold value also depends on the shape of the energy density profile \cite{Musco3,Escriva3,Musco4}.

Using the equation of state, we calculate the threshold for collapse following the method of \cite{Musco2,Byrnes}. To do this, we track the equation of state at the horizon entry time of the perturbation and the time at which the overdensity begins to collapse. These times correspond to the minimum and maximum peak formation times, which can be related to minimum and maximum peak masses using \cref{horizonmass}. This can be intuitively understood as the horizon mass growing with time. 

\begin{figure}[!ht]
    \centering
    \includegraphics[width=\linewidth]{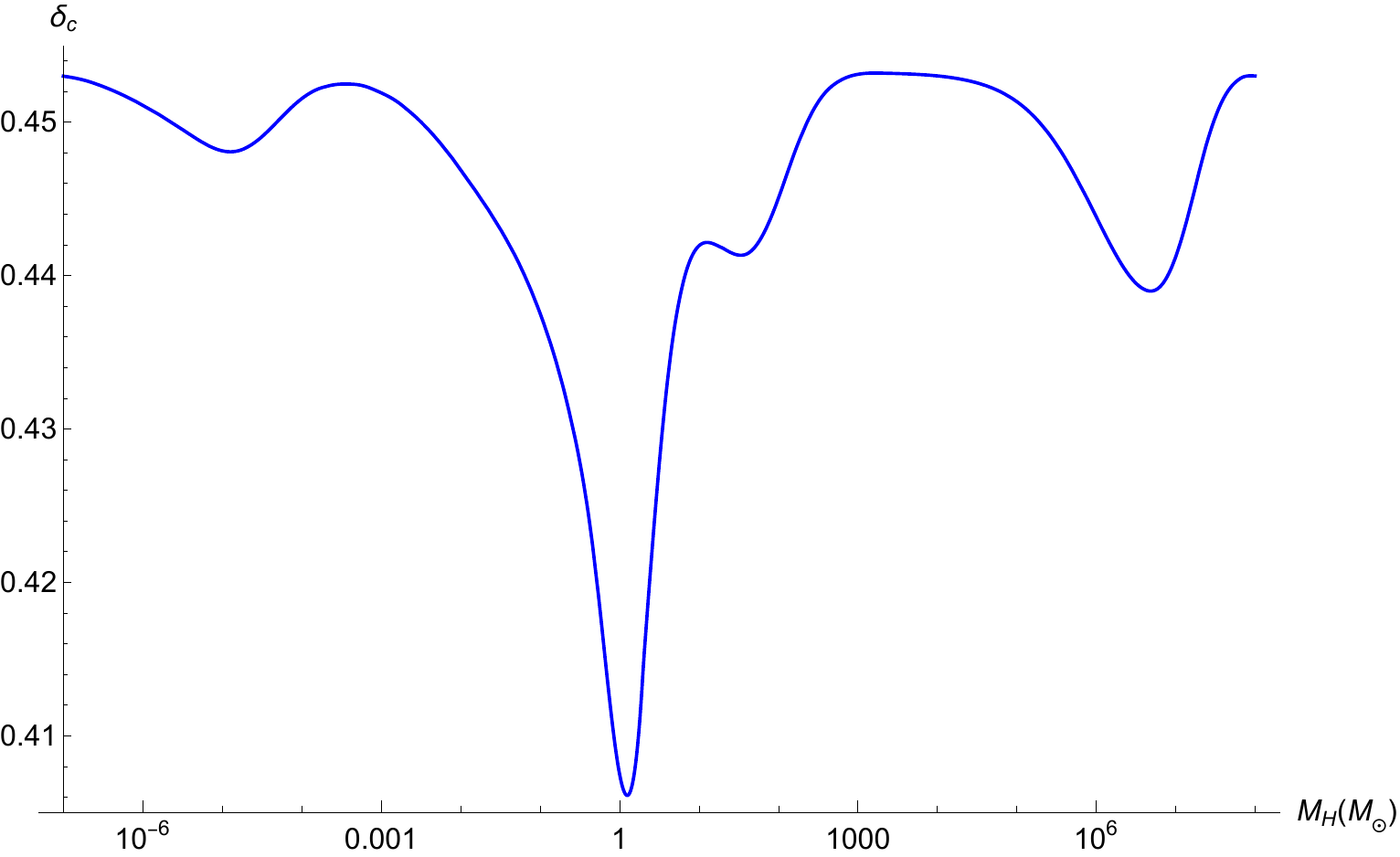}
    \caption{The threshold for the collapse of PBHs, as a function of the horizon mass of the universe. Here we take the radiation domination value of the threshold as 0.453, which has been found numerically \cite{Musco2}. Clearly, there are three departures from this value, corresponding to the three PTs described in the previous subsection.}
    \label{thresholdplot}
\end{figure}

It is also useful to track time-averaged and logarithmic time-averaged values from the equation of state. These values are both similar, suggesting an insensitivity to our method of averaging. It should also be stated that the averaged values are likely to be closer to the true answer, due to the equation of state varying between horizon entry and overdensity collapse. We plot $\delta_c(M_H)$ in \cref{thresholdplot}, obtained using the time-averaging method.

When calculating many quantities involved with PBHs there are typically many layers of detail, both theoretically and numerically. For example, in the above derivation of $\delta_c$ from $\omega$. As noted in \cite{Franciolini}, for a more complete calculation we should include the effect of both the sound speed squared and the gravitational field potential on the dynamics of PBH collapse. In this example, we find that although these inclusions shift each drop in $\delta_c$ to slightly higher masses, the shape remains similar. Throughout the paper, we have tried to be careful in selecting methods that lead to results that are insensitive to increases in complexity, which we exemplify in  \cref{Appendix-extension}.

\newpage

\section{PBH mass distribution}\label{section 3}

The abundance of PBHs is usually stated in terms of the mass fraction of the universe that collapses to form PBHs at the time of formation or the fraction of CDM comprised of PBHs. These, in themselves, are interesting quantities to study, with a plethora of techniques used in calculations throughout the literature. Moreover, knowing the PBH abundance is pivotal to discussions about GWs from black hole mergers, constraints on PBHs, and gravitational wave backgrounds to name a few.

\subsection{The curvature power spectrum}

As previously mentioned, there are several challenges with having an extended, nearly flat primordial power spectrum if we wish to motivate PBH abundance enhancement due to the drops in $\omega(T)$ \cite{Serpico}. We aim to test these conclusions for different shapes of power spectra, of which we focus on two commonly used in the literature. The two we focus on are different parametrisations of peaks with tuneable positions and widths, which offer a more realistic situation than a flat peak with sharp cutoffs; generating a 6--7 orders of magnitude increase in power at a single scale is easier for a model than creating the same increase in power over a large range of scales.

The first peaked parametrisation we consider is the lognormal power spectrum, defined as
\begin{equation}\label{lognorm}
    \mathcal{P}_\zeta(k) = A\frac{1}{\sqrt{2\pi}\Delta}\text{exp}\left(-\frac{\text{ln}^2(k/k_p)}{2\Delta^2}\right),
\end{equation}
where \(\Delta\) is the width of the power spectrum, \(k_p\) is the peak scale of the power spectrum and the normalisation is chosen such that the variance with respect to $\zeta$ is equal to $A$. We recover the Dirac delta function for $\Delta\rightarrow0$.

Secondly, we focus on the broken power law power (BPL) spectrum, defined as
\begin{equation}\label{BPL}
    \mathcal{P}_\zeta(k) = A\frac{(\alpha+\tau)^\lambda}{\left[\tau(k/k_p)^{-\alpha/\lambda}+\alpha(k/k_p)^{\tau/\lambda}\right]^\lambda},
\end{equation}
where $\alpha$ describes the growth, $\tau$ describes the decay and $\lambda$ the width of the spectrum at the scale $k_p$. This parametrisation arises naturally from USR models \cite{Atal:2018neu}. In both these examples, $A$ is the power spectrum amplitude, however, for the BPL $A$ does not correspond to the variance.

In the literature, it is common to see BPL shape parameters described with different notation. We have chosen our notation so as not to confuse later discussion, and have been consistent throughout the paper.

\subsection{Variance of perturbations}\label{variancesection}

Once we have chosen a curvature power spectrum we can define the smoothed density power spectrum during radiation domination as
\begin{equation}\label{PdeltaR}
    \mathcal{P}_{\delta_R} (k) = \frac{16}{81}(kR)^4W^2(k,R)\mathcal{P}_\zeta(k),
\end{equation}
where \(W(k,R)\) is a window function used to relate $\mathcal{P}_\zeta(k)$ in Fourier space to the probability distribution function of $\delta$ in real space \cite{Ando,Kalaja:2019uju}. The 16/81 factor depends on the equation of state which is non-constant during a given PT, making this equation approximate. We also choose to omit a transfer function \(T(k,R)\) because it's order one at horizon entry. It has, however, recently been argued that a non-linear transfer function is needed to be consistent in threshold and variance calculations \cite{DeLuca}. Moreover, the renormalisation procedure \cite{franciolini2} must also be applied on super-horizon scales.

As discussed in \cite{Young}, we must use the same smoothing function when calculating the PBH formation criterion. In this work, we mainly considered a Gaussian window function modified by a factor of 2 in the exponent \cite{Young} defined as
\begin{equation}\label{GaussianWindow}
    W(k,R)=\text{exp}\left(-\frac{(kR)^2}{4}\right).
\end{equation}
We also consider the Fourier transform of the top-hat smoothing function, given by
\begin{equation}\label{RTHwindow}
    \Tilde{W}(k,R)= 3\frac{\text{sin}(kR)-kR\,\text{cos}(kR)}{(kR)^3}.
\end{equation}
We can then use $\mathcal{P}_{\delta_R}(k)$ to calculate the momenta of the distribution, given as
\begin{equation}\label{Variance}
    \sigma^2_i(R)=\int_{0}^{\infty} \frac{\text{d}k}{k} k^{2i} \mathcal{P}_{\delta_R},
\end{equation}
where $i$ is a non-negative integer.

We note here an important point regarding both of the considered power spectra. Upon increasing the respective width parameters, the variance of perturbations with respect to the density contrast (\cref{Variance}, with $i$=0) increases (decreases) for lognormal (BPL) power spectra. This will be an important distinction when discussing PBH constraints later in this paper.

\subsection{Press–Schechter formalism}\label{PSformalism}

The mass fraction of the universe that collapses to form PBHs at the time of formation, $\beta$, is a possible way of describing PBH abundances. This can loosely be thought of as the effect the curvature power spectrum and the equation of state in the early universe have on each other. Typically, this is calculated using one of two formalisms: Press--Schechter theory \cite{Press} or peaks theory \cite{Bardeen}. The focus of this subsection and the vast majority of the paper is the former. We note that the power spectrum amplitude required to generate a given PBH abundance is insensitive to the calculation technique \cite{Gow}. 

In the context of PBHs, Press-Schechter theory says that density perturbations with a peak value greater than the threshold value, $\delta_c$, will collapse to form a black hole. The abundance is thus given as
\begin{equation}\label{PressSchectbeta}
    \beta=2\int_{\delta_c}^{\infty}\text{d}\delta P(\delta),
\end{equation}
where the factor of 2 is the Press-Schechter factor. If we assume that perturbations follow a Gaussian distribution then the Gaussian probability density, $P$($\delta$), is given as
\begin{equation}\label{GaussProbablity}
    P(\delta)=\frac{1}{\sqrt{2\pi \sigma_0^2}}\text{exp}\left( -\frac{\delta^2}{2\sigma_0^2}  \right).
\end{equation}
If we further assume that PBHs form with exactly the horizon mass we can then approximate the abundance as
\begin{equation}
    \beta=\text{erfc}\left( \frac{\delta_c}{\sqrt{2\sigma_0^2}} \right), \label{AbundanceBeta}
\end{equation}
where erfc is the complementary error function. Finally, we can compute the mass distribution of PBHs stated in terms of the fraction of CDM made up of PBHs of a given mass $M$, given as
\begin{equation}\label{fPBH}
    f(M_{\text{PBH}})=\frac{1}{\Omega_{\text{CDM}}}\frac{\text{d}\Omega_{\text{PBH}}}{\text{dln}M},
\end{equation}
where $\Omega_{\text{PBH}}$ is the total abundance of PBHs relative to the critical density and $\Omega_{\text{CDM}}=0.245$ \cite{aghanim}. This equation is simplified if $\beta$ is constant when $\delta_c$ is constant, which is the case if we assume that the density power spectrum at horizon-entry is scale-invariant over a relevant but limited range of scales. If we further assume that black holes form with mass $M$=$M_H$, we arrive at
\begin{equation}\label{fPBH1}
    f(M_{\text{PBH}}) = \frac{2.4}{f_{\text{PBH}}} \beta(M_\text{PBH}) \left( \frac{M_{\text{PBH}}}{M_{\text{eq}}}\right) ^{-1/2},
\end{equation}
where the factor 2.4 comes from 2(1+$\Omega_\text{b}/\Omega_{\text{CDM}})$, with $\Omega_{\text{b}}=0.0456$ \cite{aghanim} and $M_{\text{eq}}$ is the horizon mass at the time of matter-radiation equality, $M_{\text{eq}} \approx2.8\times10^{17}$ $M_\odot$.

In a more realistic calculation, we should take into consideration the effect critical collapse has on the PBH mass spectrum. The previous calculation can thus be extended to account for PBHs not forming with exactly the horizon mass. It was found that the PBH mass follows a scaling relation \cite{Niemeyer:1997mt,Musco2}, described by the critical collapse equation
\begin{equation}\label{MassScaling}
    M_{\text{PBH}} = \kappa M_H (\delta-\delta_c)^{\gamma},
\end{equation}
where $M_H$ is the horizon mass at horizon re-entry, $\kappa\simeq0.36$, $\gamma\simeq3.3$ and $\delta_c\simeq0.453$ are found numerically assuming radiation domination. We arrive at the amended expression
\begin{equation}\label{betaMassScaling}
    \beta=2\int_{\delta_c}^{\infty}\text{d}\delta\frac{M_{\text{PBH}}}{M_H}P(\delta)=2\int_{\delta_c}^{\infty}\text{d}\delta\kappa(\delta-\delta_c)^{\gamma} P(\delta).
\end{equation}
Defining $\mu\equiv\frac{M_{\text{PBH}}}{\kappa M_H}$ and $\frac{\text{d}\delta}{\text{d}\text{ln}M_{\text{PBH}}}=\frac{1}{\gamma}\mu^{1/\gamma}$, we obtain the PBH mass distribution today 
\begin{equation}{\label{fPBHMassScaling}}
  \begin{split}
   f(M_{\text{PBH}})= & \frac{1}{\Omega_{\text{CDM}}}\int_{-\infty}^{\infty} \text{d}\text{ln}M_H \frac{2}{\sqrt{2\pi\sigma_0^2(M_H)}}\text{exp}\left[ -\frac{(\mu^{1/\gamma}+\delta_c(M_H))^2}{2\sigma_0^2(M_H)} \right]\\
     & \qquad \qquad \qquad \qquad \qquad \qquad \times\frac{M_{\text{PBH}}}{\gamma M_H} \mu^{1/\gamma}\sqrt{\frac{M_{\text{eq}}}{M_H}}.
   \end{split}
\end{equation}
\subsection{Phase transition impacts on PBH abundances}\label{ratiosection}

We would like to quantify a given PT's impact on PBH abundances. We first calculate the scale corresponding to the peak of $f(M_{\text{PBH}})$ assuming radiation domination, which we label $k_{f_{\rm peak}}$. This scale depends on power spectrum parameters, and is not equal to the peak scale of $\mathcal{P}_\zeta$. We then consider the scales at which the threshold for collapse is at a minimum, for each SM PT. These scales, which we label $k_{PT}$, are fixed by SM physics. On the other hand, the value of $f(M_{\text{PBH}})$ at these scales clearly depends on the power spectrum shape as well as its amplitude.

Considering the collapse threshold shown in \cref{thresholdplot}, we calculate the ratio of $f(M_{\text{PBH}})$ at $k_{PT}$ versus $k_{f_{\rm peak}}$. We do this analysis for the parameter space of each of the considered power spectra, normalising the power spectrum amplitude $A$ such that $f_\text{PBH}=1$, which gives us a better understanding of the effect of varying each parameter. In reality, this value of $f_\text{PBH}$ is constrained at the QCD scale to be $\mathcal{O}$(10$^{-3}$) \cite{carcasona}. Notably, we have checked the ratio $f_{\rm PT}/f_{\rm peak}$ is not very sensitive to $f_\text{PBH}$.

If this ratio is unity, the value of $f(M_{\text{PBH}})$ is the same at $k_{f_{\rm peak}}$ and $k_{PT}$. Of course, this will occur if the scales are the same (i.e. $k_{f_{\rm peak}}=k_{PT}$). We argue that in this case, despite a given PT increasing $f(M_{\text{PBH}})$, there would have to be some level of fine-tuning in a given model to realise the scenario. Of interest are the parameter values that lead to the ratio being greater than one, signifying a significant change in the shape of our abundance curve which will now have its highest value at the same scale as the PT. In these scenarios, we can say that a given PT has impacted the abundance of PBHs.

\begin{figure}[!ht]
    \centering
\includegraphics[width=\linewidth]{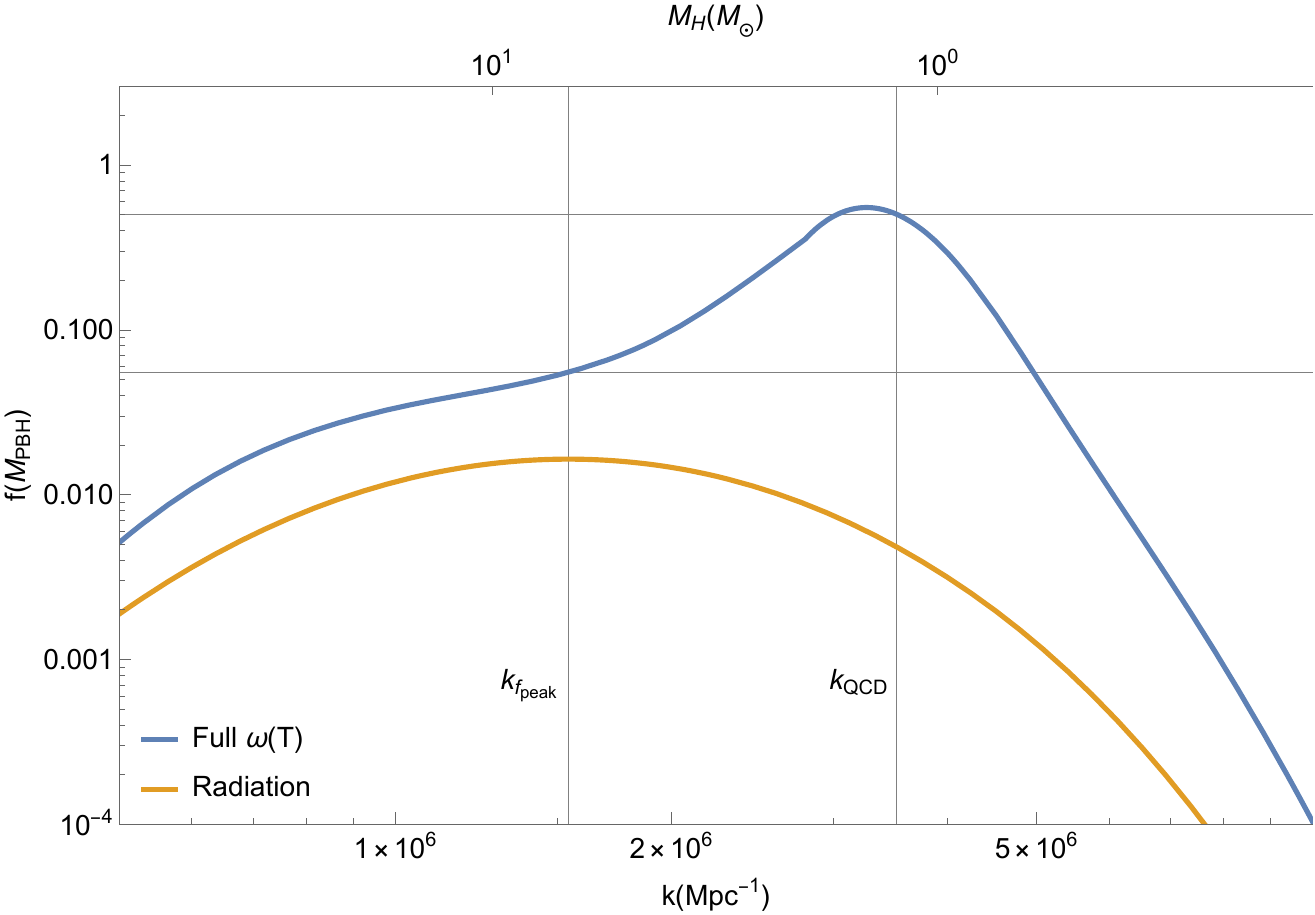}
    \caption{We exemplify our method of calculating the impact of the QCD PT on the PBH abundance. We show the abundance curves assuming radiation domination and considering the full equation of state, for a lognormal power spectrum with $k_p=2\times10^6$ and $\Delta=2.5$. The vertical grid lines show the peak scale of the abundance curve assuming pure radiation and the QCD scale. The horizontal grid lines are the corresponding values of the full $f(M_{\text{PBH}})$ evaluated at these scales. In this example, we recover a ratio of $\approx9.14$.}
    \label{ratioexplained}
\end{figure}

Initially, we focus on a lognormal $\mathcal{P}_\zeta$ peaked at scales around the QCD crossover. The biggest drop in relativistic degrees of freedom occurs on these scales so we expect the biggest change in the shape of $f(M_{\text{PBH}})$ to be around this QCD scale, upon making $\delta_c$ a function of $M_H$. In \cref{ratioexplained}, we exemplify our process of calculating $f_{\rm PT}/f_{\rm peak}$, for lognormal parameters $k_p=2\times10^6$ and $\Delta=2.5$, recovering a value of $\approx9$. This value is relatively large, so we would say that the QCD PT has had a large impact on the PBH abundance curve, which can be seen from this figure. 

Interestingly, we see in the figure that the peak in the full $f(M_{\text{PBH}})$ does not quite occur at the \say{QCD scale}. Of course, the QCD PT isn't instantaneous and many scales are affected by it. As previously stated, we defined this to be the scale of the minimum in the collapse threshold during the QCD PT. The reason for this result can be seen by looking at \cref{AbundanceBeta}. Specifically, the abundance is dictated by the ratio within the complementary error function. In this example, the standard deviation of perturbations has increased quicker than the threshold (for $k\lesssim k_\text{QCD}$), resulting in $\delta_c/\sqrt{2\sigma_0^2}$ decreasing which corresponds to a larger abundance. We find that this property becomes more evident as $k_p$ decreases. 

We considered calculating the ratio at the peak scale in the full $f(M_{\text{PBH}})$, instead of at the QCD scale. However, regardless of at which scale you evaluate $f(M_{\text{PBH}})$ in the numerator, the ratio remains relatively similar. We simply point out that, as the difference in $k_p$ and $k_\text{QCD}$ increases, our ratio values increasingly underestimate this new ratio. 

\begin{figure}[!ht]
    \centering
    \includegraphics[width=\linewidth]{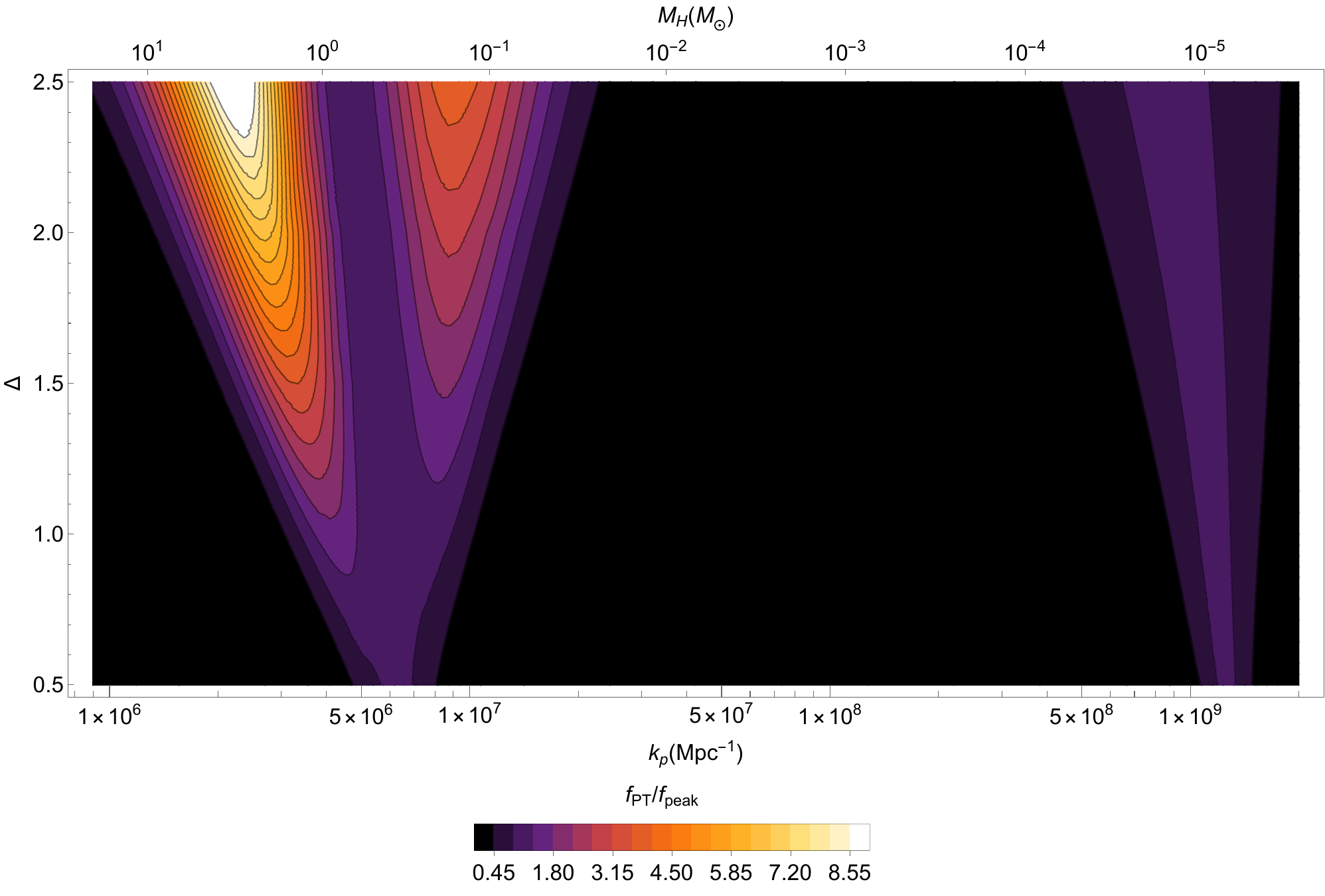}
    \caption{The ratio of the full $f(M_{\text{PBH}})$ evaluated at PT scales versus the peak scale in the abundance curve assuming radiation domination.  We show the results assuming a lognormal power spectrum, over the QCD and EW PTs. This gives us some intuition about the impact of a PT on the shape of $f(M_{\text{PBH}})$ as the $\mathcal{P}$$_\zeta$ parameters are varied. At the EW scale, we see the maximum of this ratio equaling unity.}\label{lognormratio}
\end{figure}

In \cref{lognormratio}, we show the value of the ratio, $f_{\rm PT}/f_{\rm peak}$, assuming a lognormal peak. We can see from the figure that, for a given peak scale in $\mathcal{P}_\zeta$, increasing the width will lead to an increase in the ratio. This is to be expected, for broader spectra scales further from the peak will have more power. As previously mentioned, the PBH abundance is very sensitive to this increase. By widening our spectra we are thus enhancing the impact of the drop in $\delta_c$ on the shape of our abundance curve.

As well as the QCD PT, we investigated the EW PT in this context. These results are also plotted in \cref{lognormratio} where we can see that none of the [$k_p$, $\Delta$] parameter space leads to a ratio $f_\text{EW}$/$f_{\rm peak}$ greater than unity. This doesn't come as much of a surprise, despite the exponential sensitivity of abundances to $\delta_c$($\omega$) we only see a drop of $\sim$ 1$\%$ in this parameter (as opposed to $\sim$ 10$\%$ during the QCD PT). In \cite{cosmicconundra}, two planetary-mass microlensing events \cite{Niikura,Niikura1} were argued to possibly be explained by PBHs. The motivation for this was the softening in $\omega$ at the EW scale producing a peak in their abundance curve. However, our results suggest difficulty motivating PBHs at this precise scale, assuming a lognormal power spectrum.

Despite having the ability to choose lognormal parameter values which lead to a large enhancement on the impact of the QCD PT on PBH abundances, many parameter values are ruled out by observational constraints. These will be discussed in \cref{section 4}.

\begin{figure}[!ht]
    \centering
    \includegraphics[width=0.8\linewidth]{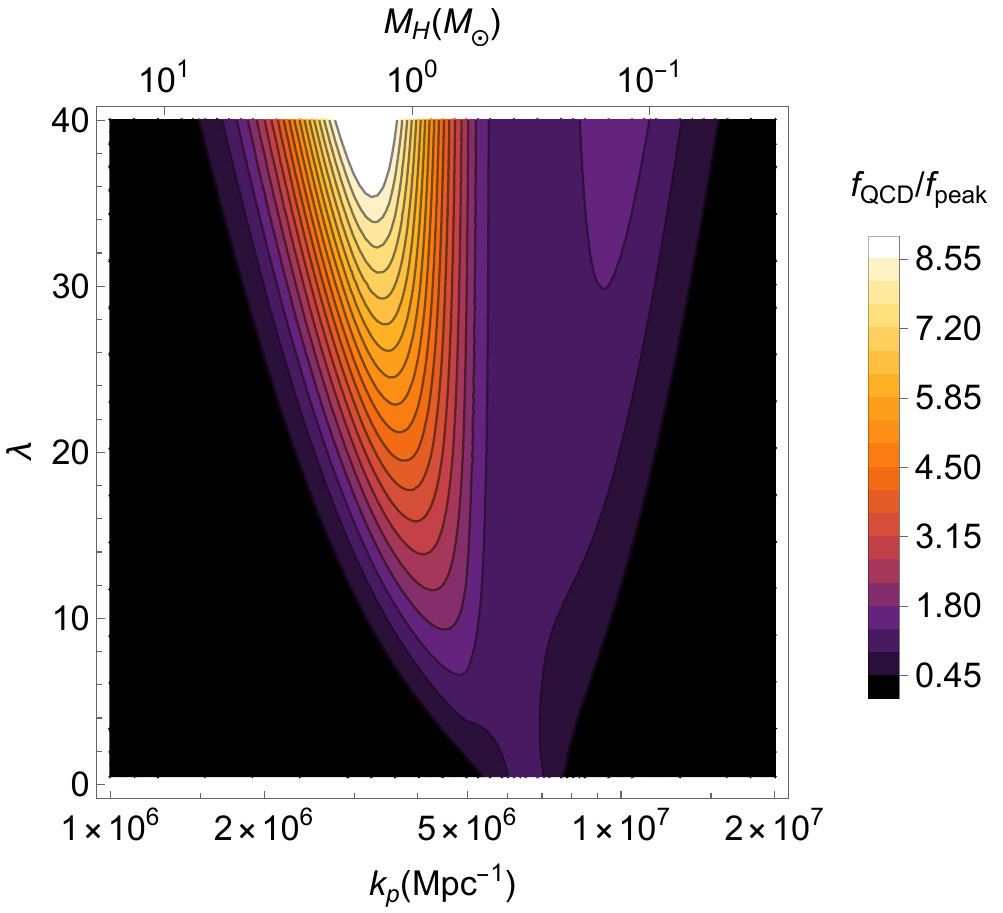}
    \caption{The ratio of the full $f(M_{\text{PBH}})$ evaluated at the QCD scale versus the peak scale in the abundance curve assuming radiation domination, for a BPL power spectrum. In this plot, we vary the width parameter, $\lambda$, whilst keeping $\alpha=4$ and $\tau=3$. We can see several similarities to \cref{lognormratio}.}
    \label{BPLratio}
\end{figure}

Lastly, we investigated the effect of the QCD PT on the PBH abundance curve assuming a broken power law power spectrum. To make a more direct comparison with \cref{lognormratio}, we focus on the width parameter of the broken power law, $\lambda$. Our results are plotted in \cref{BPLratio}, where we have set shape parameters $\alpha=4$ and $\tau=3$. Plotting up to $\lambda=40$, we reproduce a similar plot to the lognormal, with some interesting distinctions. It would be informative to investigate whether such a large value for $\lambda$ is natural from a model-building perspective. One could also perform a log-likelihood analysis on a relevant dataset as was done in \cite{Franciolini1} (although only up to $\lambda=10$), to reveal whether this value of $\lambda$ was favorable. However, these discussions are beyond the scope of this paper.

The figure exemplifies how important the cut-offs of a power spectrum are when calculating $f_\text{PT}$/$f_{\rm peak}$. For $k_p \lesssim 6\times10^6$ Mpc$^{-1}$ (which is the scale at which $f_{\rm peak}$=$f_\text{QCD}$), $\tau$ is the relevant quantity. For $k_p$ greater than this scale, $\alpha$ is the relevant quantity. Decreasing the steepness of the slope by a value of 1, we see a large increase in the ratio. We point out that this effect isn't solely due to the slope. As can also be seen in the lognormal case (which has the same slope on either side of the peak), the ratio is typically larger for lower values of $k_p$. This is due to pions ($\pi^{\pm,0}$) and muons ($\mu$) decoupling shortly after the QCD PT, leaving an additional drop in $\delta_c$, which can be seen in \cref{thresholdplot}.

\section{PBH constraints}\label{section 4}

Observational constraints have been placed on PBH abundances over a large range of masses (See \cite{Carr} for a review). The majority of these constraints come from direct observations where, if there were a certain abundance of PBHs at a given mass, we would expect to see their effect on a given observation. On the other hand, see \cite{Carr3} for hints of detections from various observations. As well as direct observational constraints, PBHs also face indirect constraints. These are associated with the primordial curvature power spectrum. Specifically, the large increase in power required for the formation of PBHs in the typical scenario. In this section, we discuss both direct and indirect constraints, followed by a comparison of their respective constraints on power spectrum parameters.   

\subsection{Direct constraints}

In the standard cosmological picture, primordial black holes with initial masses greater than $m\sim10^{15}$ g wouldn't have completely evaporated by today. Thus, direct constraints on PBH abundances of lower masses come from the effect of their Hawking evaporation \cite{Hawking:1975vcx}. For example, the Hawking evaporation of PBHs can affect element abundances \cite{1977SvAL....3..110Z,Keith:2020jww}, $\gamma$-ray backgrounds \cite{1976ApJ...206....1P,Carr:2016hva} and cosmic rays \cite{1991ApJ...371..447M,Ziparo_2022}. PBHs with masses in the range $10^{17}\text{g}\lesssim m \lesssim10^{22}\text{g}$ are unconstrained, meaning they can account for the totality of dark matter. It should be stated that this \say{asteroid-mass window}, as well as each of the direct observational constraints, depend on the PBH mass-function one assumes \cite{Gorton_2024}. Interestingly, in higher dimensions the evaporation bound is weakened \cite{Guedens}, leaving a larger range in which PBHs can account for the totality of dark matter.  

For PBHs with masses $10^{22} {\rm g}\lesssim m$, there are also a large number of direct constraints. These include microlensing \cite{Niikura1} and dynamical \cite{1999ApJ...516..195C} constraints. To constrain $\mathcal{P_\zeta}$ from the direct constraints on PBH abundance, we first use Fig.~18 from \cite{Carr} which shows the combined constraints on the mass fraction $\beta$ with common parameters normalised out. We use Eq.~8 of the same paper to relate this parameter to $\beta$ as follows
\begin{equation}\label{betaConstraints}
\beta'(M)\equiv\gamma_{\text{eff}}^{1/2}\left(\frac{g_{\ast f}}{106.75}\right)^{-1/4}\left( \frac{h}{0.67}\right)^{-2}\beta(M),
\end{equation}
where $h\approx$ 0.67 \cite{aghanim} is the dimensionless Hubble parameter and $g_{\ast f}$ is the standard model relativistic degrees of freedom at the epoch of PBH formation. We set $\gamma_{\text{eff}}$ equal to unity, however, the precise value is an ongoing debate \cite{Kozaczuk}. 

We then use the constraints on the mass fraction and substitute \cref{AbundanceBeta} into \cref{betaConstraints}, and rearrange for the power spectrum amplitude. This results in the amplitude constraint plot being a function of power spectrum parameters. We note that the formation criterion for PBHs depends on the choice of smoothing function \cite{Young,Young3}. As such, using the modified Gaussian window function in our abundance calculation, we set $\delta_c$=0.25. Due to the exponential sensitivity of PBH abundance to the threshold, the correct choice of $\delta_c$ is very important to the discussion of power spectrum constraints. 

We account for the non-linearities between the density contrast and curvature perturbation \cite{Harada} following the arguments of \cite{Gow}. We thus multiply our constraint line by a factor of 1.98. In \cref{NonlinearComp} we compare the full non-linear calculation (detailed in \cref{Appendix-NL}) to the simpler approximation. We see excellent agreement between the two, justifying the simplification. 

\begin{figure}[!ht]
    \centering
    \includegraphics[width=\linewidth]{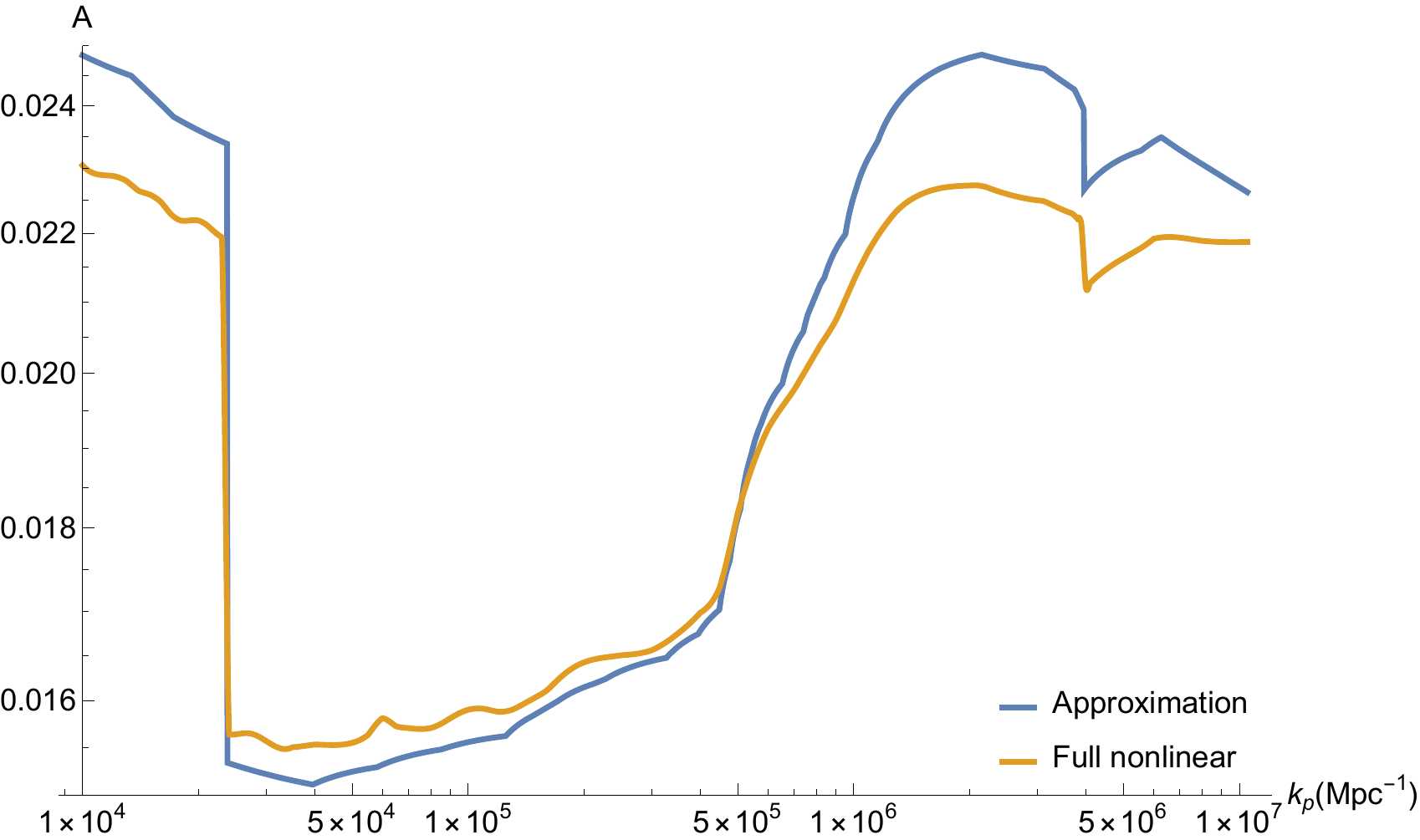}
    \caption{The direct observational constraints of PBHs on the primordial power spectrum. The sudden drop comes from CMB accretion constraints \cite{Agius:2024ecw}. We compare the full calculation including non-linearities to the approximation made in \cite{Gow}, which we use for the remainder of the paper. We use a lognormal power spectrum with $\Delta$=1. We see excellent agreement between this approximation and the more computationally expensive full calculation.}
    \label{NonlinearComp}
\end{figure}

As previously mentioned, it was found that the power spectrum amplitude required to generate a given value of $f_{\text{PBH}}$ is insensitive to whether we calculate abundances via Press--Schechter or peaks theory in Ref.~\cite{Gow} (but see \cite{Iovino:2024uxp} for a different view). If we instead fixed the power spectrum amplitude and then calculated the abundance, we expect a 1--2 order of magnitude difference for each technique \cite{Young:2014ana,Wang:2021kbh}. This difference is important for some purposes, however, for this paper, we are interested in constraints on power spectrum amplitude. Thus, it is more important to us that the power spectrum amplitude remains similar regardless of the abundance calculation technique. The large difference in PBH abundance versus small difference in power spectrum amplitude constraints is due to the exponential sensitivity of the former to the latter. This means that, if we were to use peaks theory, we would expect similar limits on the power spectrum from direct PBH constraints.

\subsection{CMB \texorpdfstring{$\mu$}{μ}-constraints}

Spectral distortions quantify deviations from the CMB's black-body temperature distribution. If there were an increase in power on small scales the energy stored in density perturbations would be dissipated through photon diffusion, via Silk damping. On scales relevant to $\mu$-distortions (the most relevant for our analysis), Bremsstrahlung and double Compton scattering are inefficient due to the expansion of the universe. These inefficiencies result in photon distributions on different scales being described by different blackbodies \cite{Chluba:2015bqa}.

The spectral distortions place one of the tightest current constraints on the primordial curvature power spectrum. During the $\mu$-era, due to no detection of spectral distortions, the power spectrum amplitude must be lower than 10$^{-4}$, meaning PBH formation at this scale is completely ruled out in the standard picture of Gaussian curvature perturbations \cite{Chluba:2012we}. Interestingly, this scale corresponds to PBHs of supermassive black hole masses and $e^-$$e^+$ annihilation. Adding non-Gaussianity with positive skewness is not enough to stop $\mu$-constraints from being significantly tighter than the PBH constraints unless very extreme and completely non-perturbative non-Gaussianity is considered \cite{Nakama,hooper,Sharma:2024img,Byrnes:2024vjt,Iovino:2024uxp}. 

To quantify how much additional power is needed to cause observable spectral distortions we must calculate the final $\mu$-distortions induced by scalar perturbations, which can be approximated as \cite{Chluba:2015bqa}
\begin{equation}\label{muapprox}
     \mu \approx \int_{k_{\text{min}}}^{\infty} \frac{\text{d}k}{k} \mathcal{P}_\zeta(k) W_\mu(k),
\end{equation}
with $k$-space window functions of the form 
\begin{equation}\label{muwindow}
    W_\mu(k)\approx 2.27\left[ \text{exp} \left( -\left[\frac{\hat{k}}{1360}\right]^2 \middle/ \left[ 1 + \left[ \frac{\hat{k}}{260} \right]^{0.3} + \frac{\hat{k}}{340} \right] \right) -\text{exp}\left( -\left[\frac{\hat{k}}{32}\right]^2\right)\right],
\end{equation}
where $\hat{k}\equiv k$/1 Mpc$^{-1}$ and $k_{\text{min}}$ $\simeq$ 1 Mpc$^{-1}$. We can compare the total induced $\mu$-distortions with observations, resulting in a power spectrum-dependant function of scale $k$. The strongest current observational constraint on $\mu$-distortions comes from the FIRAS instrument on the COBE satellite. A recent reanalysis of FIRAS data suggests $\mu$-distortions to be smaller than 4.7$\times$10$^{-5}$ \cite{Bianchini}, which is 2 times tighter than the original analysis \cite{Fixsen}.

\subsection{Pulsar timing arrays}\label{PTAsection}

On smaller scales than $\mu$-distortion constraints, the primordial curvature power spectrum is constrained by PTAs. An increase in power at a given scale will result in large first-order scalar perturbations at that scale. In cosmological perturbation theory at second order, scalar and tensor perturbations are coupled meaning this increase in power can be responsible for a stochastic gravitational wave background (see \cite{Domenech1} for a review).

Recently, PTA collaborations NANOGrav \cite{NANOGrav,NANOGrav1}, EPTA \cite{EPTA,EPTA1,EPTA2}, PPTA \cite{PPTA,PPTA1,PPTA2} and CPTA \cite{CPTA} have released data showing evidence for a stochastic gravitational wave background. There is already much literature investigating PTA results and $\mathcal{P}_\zeta$ \cite{PTApz,PTApz1,PTApz2,PTApz3}, equation of state \cite{Zhu:2023gmx,Liu:2023pau}, as well as PBH studies \cite{Franciolini1,PTApbh1,PTApbh2,PTApbh3,PTApbh4,PTApbh5,PTApbh6,PTApbh7,Wang_2024,Huang:2023mwy}. In addition to scalar-induced GWs there are many other cosmological explanations for this evidence \cite{NANOnew,Ye:2023tpz}, such as phase transitions \cite{PTApt,PTApt1,PTApt2,PTApt3,PTApt4,PTApt5,PTApt6,He:2023ado} and topological defects \cite{PTAtd,PTAtd1,PTAtd2,PTAtd3,PTAtd4,PTAtd5}. There are also astrophysical explanations for the signal \cite{PTAastro,PTAastro1,PTAastro2,PTAastro3,PTAastro4}, such as merging supermassive black holes. We assume that the stochastic gravitational wave background is due to something other than scalar-induced GWs, allowing us to place tighter constraints on the curvature power spectrum. 

The scalar-induced GW spectrum today, $\Omega_\text{GWB}$, from a given curvature power spectrum is given by \cite{Ananda:2006af,Baumann:2007zm}
\begin{equation}\label{GWBeq}
    \Omega_\text{GWB}(k)h^2\approx3.2\times10^{-5}\int_{0}^{1}\text{d}d\int_{0}^{\infty}\text{d}s\mathcal{T}_\text{rad}(d,s)\mathcal{P}_\zeta\left(\frac{k}{2}(s+d)\right){P}_\zeta\left(\frac{k}{2}(s-d)\right),
\end{equation}
with transfer function, $\mathcal{T}_\text{rad}(d,s)$, given in \cite{Kohri:2018awv} for example.

To calculate the scalar-induced GW spectrum we used the python package \cite{SIGWfast}, which assumes that the equation of state is constant during GW production. As discussed in \cref{SM PTs}, this is not the case for the early universe. A full calculation is beyond the scope of this paper, however, see \cref{constraintMH} for more discussion. 

Then, we find the limiting amplitude of the given power spectrum such that PTA data may constrain the scalar-induced GW spectrum. We use the all-sky GW strain 95$\%$ upper limit data from NANOGrav15 \cite{NANOdata}, which we convert to a limit on primordial GWB energy density through the following equation
\begin{equation}\label{GWB}
    \Omega_{\text{GWB}}(f)h^2=\frac{2\pi^2}{3H_0^2}f^2h_c^2(f),
\end{equation}
where $H_0$=100 $h$ km s$^{-1}$ Mpc$^{-1}$, and $h_c$ is the GW strain. We convert this to units of wavelength via the following equation
\begin{equation}\label{frequencytowavelength}
    f=1.6\text{nHz}\left(\frac{k}{10^6 \text{Mpc}^{-1}}\right).
\end{equation}
Finally, we arrive at our formula for constraining the power spectrum amplitude using PTA data
\begin{equation}\label{PTAconstraintEQ}
    A_{\text{constraint}}=\text{Min}\left(\sqrt{\frac{\Omega_{\text{GWB,NG}}(k)h^2}{\Omega_{\text{GWB,signal}}(k,\mathcal{P}_\zeta)h^2}}\right),
\end{equation}
where $\Omega_{\text{GWB,NG}}(k)$ is the constrained GWB value of NANOGrav and $\Omega_{\text{GWB,signal}}(k,\mathcal{P}_\zeta)$ is the GWB we calculate using \cref{GWBeq}.

The strain amplitude upper limit data was calculated in \cite{NANOdata} in two ways, using a uniform and log-uniform prior. Assuming a uniform prior corresponds to each value in a model parameter's, $\theta$, distribution to be equally likely. A log-uniform prior corresponds to a uniform distribution in ln($\theta$). A log-uniform prior is typically assumed in $\Lambda$CDM fits of $\mathcal{P}_\zeta$ amplitude \cite{Planck:2015fie,aghanim}, for example. It was shown that bounds derived with this assumed prior may be strongly sensitive to the choice of lower prior boundary \cite{Diacoumis:2018ezi}.

To place our constraints on the primordial power spectrum we chose to use the strain data assuming a uniform prior, however, see \cite{Iovino:2024uxp} for constraints assuming a log-uniform prior. The choice of prior leads to differences in power spectrum amplitude constraints of $\mathcal{O}$(2). As expected, the log-uniform prior leads to tighter constraints \cite{NANOdata}. As previously discussed, this difference of order 2 in the power spectrum amplitude leads to many orders of magnitude differences in PBH abundance, making the prior choice an important decision when calculating such constraints. Prior dependence has been discussed in many other cosmological scenarios \cite{Christodoulidis:2019hhq,Briffa:2021nxg,Patel:2024odo}.

\subsection{Constraining \texorpdfstring{$\mathcal{P}_\zeta$}{P\textunderscore{ζ}}}

\subsubsection{Lognormal peak}

\begin{figure}
    \centering
\includegraphics[width=\linewidth]{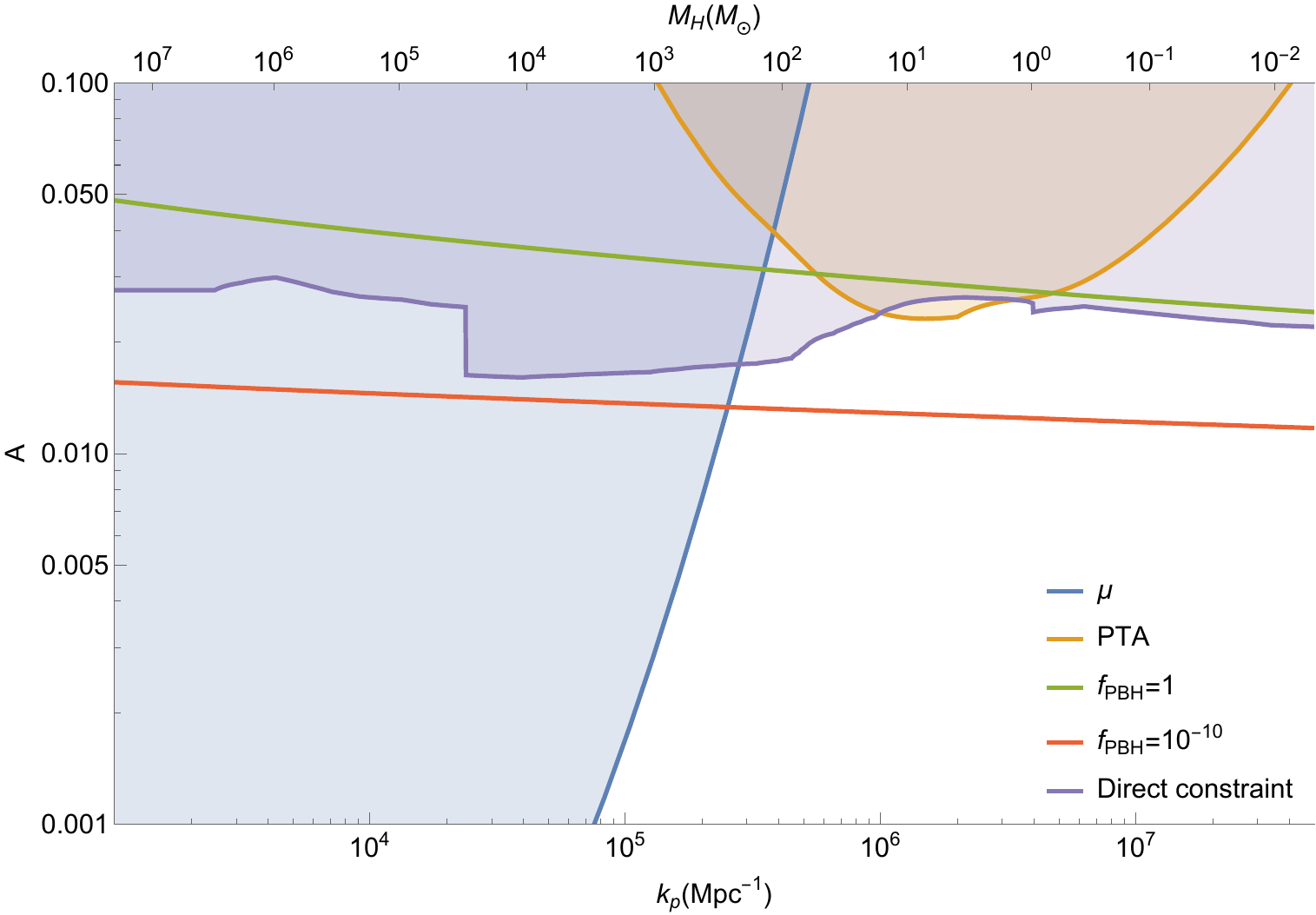}
    \caption{We show the indirect constraints on an example lognormal power spectrum, with $\Delta$=1.1. We also show the amplitude required to produce a PBH abundance $f_\text{PBH}$=1 and $f_\text{PBH}$=10$^{-10}$ and the current (direct) PBH constraint which lies in between these extreme values.}
    \label{exampleconstraint}
\end{figure}

As has been discussed, both direct and indirect observations can constrain the primordial curvature power spectrum. In \cref{exampleconstraint} we show an example plot for the constraints on the lognormal power spectrum amplitude with $\Delta=1.1$. A point above a given constraint line is ruled out by the respective constraint. We also show the values of the power spectrum amplitude leading to $f_\text{PBH}=1$ and $f_\text{PBH}=10^{-10}$. The small value of $f_{\rm PBH}=10^{-10}$ is the order of magnitude constraint we would reach if most of the dark matter consisted of WIMPs \cite{Lacki:2010zf,Adamek:2019gns,Carr:2020mqm,Boudaud:2021irr,Kadota:2021jhg,Gines:2022qzy,Eroshenko:2024dtb}.

\begin{figure}
    \centering
\includegraphics[width=\linewidth]{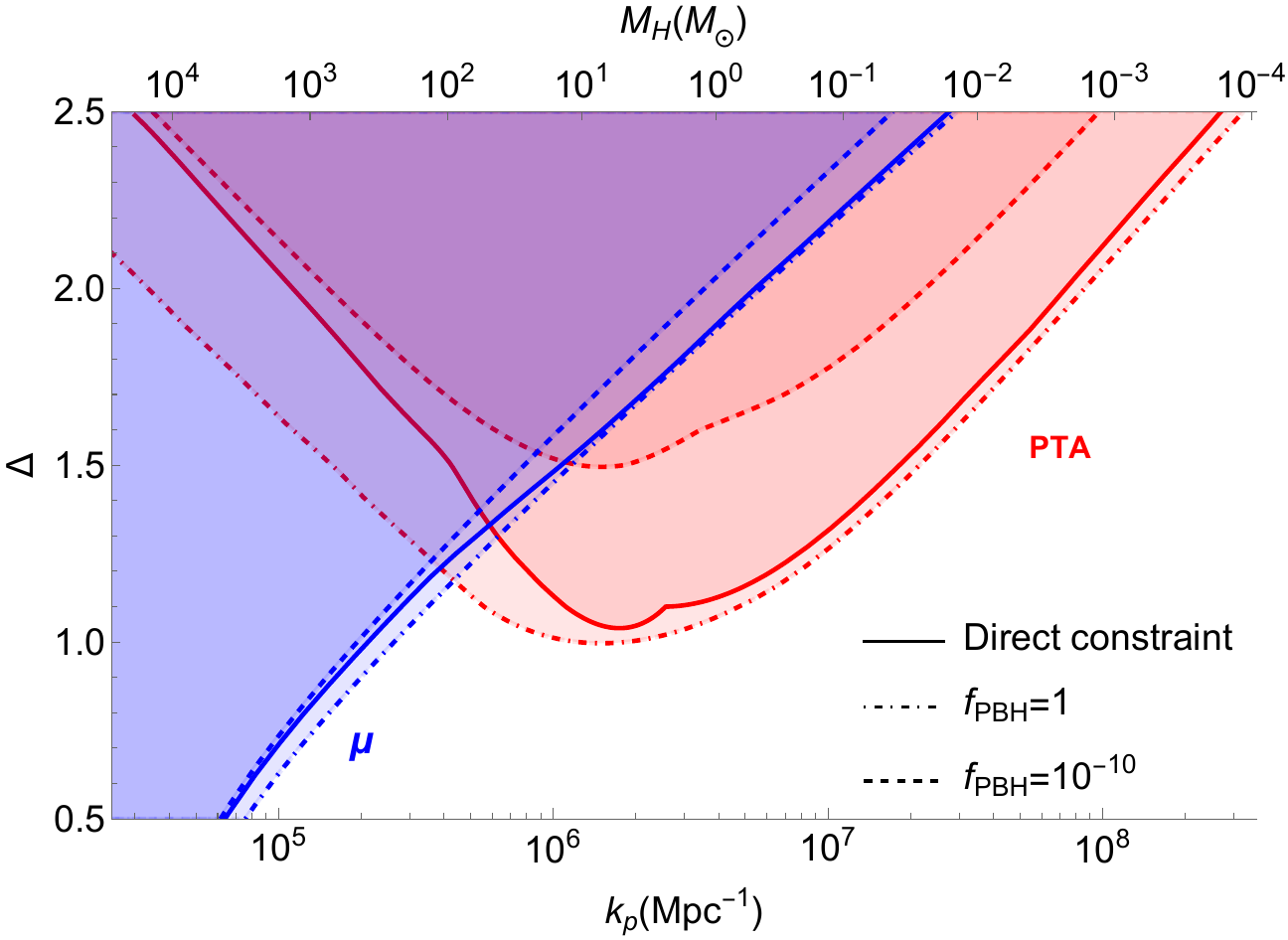}
    \caption{We show the constraints on the lognormal power spectrum from $\mu$-distortions (blue line) and PTA (red line) for different scenarios. The parameter points above these lines cannot produce PBH populations with $f_{\rm PBH}$ values above those shown in the caption.}
    \label{lognormconstraints}
\end{figure}

Using \cref{exampleconstraint}, we can identify the scales at which spectral distortion and PTA observations place tighter constraints on the power spectrum amplitude than direct PBH constraints. For the given values of $f_{\rm PBH}$ we can calculate the values of $k_p$ that are ruled out by indirect constraints. We point out that each constraint line and each $f_{\rm PBH}$ line will differ upon variation of the width of the lognormal, $\Delta$. This can be seen from each respective constraint equation, which all depend on $\mathcal{P}_\zeta$. Using all this information we can make a parameter space constraint plot, which we show in \cref{lognormconstraints}. 

Notably, $\mu$-distortion constraints on power spectrum parameters remain similar regardless of each of the scenarios considered. This shows an insensitivity to $f_\text{PBH}$ that comes from the relatively small variation in direct observational constraint lines, which is due to the exponential sensitivity of $f_\text{PBH}$ to the power spectrum amplitude. In other words, a relatively small change in $A$ will result in a relatively big change in $f_{\text{PBH}}$. With regard to the PTA parameter space constraints, we see more dependence on our PBH constraint line. This is because PTA and PBH constraints are comparably tight, resulting in a sensitivity to relatively small changes in both PTA and PBH constraint lines. We also note the coincidence of scales between PTA observations and the QCD scale ($\approx3\times10^6$ Mpc$^{-1}$), which is evident from the plot. 

We can see that a broader lognormal power spectrum leads to tighter $\mu$-distortion and PTA constraints. In particular, we find that for $\Delta\lesssim1$, PTA observations are unable to constrain the PBH abundance. In this case, it is not possible to detect QCD PT signatures on $f_\text{PBH}$ using the current PTA data. In general, the region under a given line signifies the parameter values leading to direct observational constraints on power spectrum amplitude, and thus PBH abundance, being the most competitive. If we had used GW strain data with a log-uniform prior, as discussed in \cref{PTAsection}, then each PTA constraint line in \cref{lognormconstraints} compete with direct observations to lower $\Delta$.

In \cite{cosmicconundra}, PBHs in this mass range were argued to offer solutions to several ongoing issues \cite{Mediavilla,Wyrzykowski,Kashlinsky}. Moreover, LIGO-Virgo-KAGRA detections of BHs of potentially primordial origin have also been studied in detail, including the QCD impact \cite{Franciolini,Escriva1}. We have shown that, if a lognormal $\mathcal{P}_\zeta$ doesn't peak around the QCD scale, then it must be broad for the PT to impact the overall abundance. We have also shown that this scenario is tightly constrained. We reiterate that this conclusion is drawn assuming the PTA signal is not due to scalar-induced GWs and hence we are not trying to fit PTA data.

\subsubsection{Broken power law peak}

The broken power law power spectrum could be said to have several advantages over the lognormal if we want to motivate PBHs from the drops in $\delta_c$. Firstly, instead of falling logarithmically about the peak it has sharper cut-offs. This results in scales around the peak having less of an effect on a given constraint. Secondly, with two extra parameters used to describe the BPL, we will have more freedom to control the overall power spectrum shape.

\begin{figure}
\begin{minipage}[b]{0.5\linewidth}
  \centering
  \includegraphics[width=\linewidth]{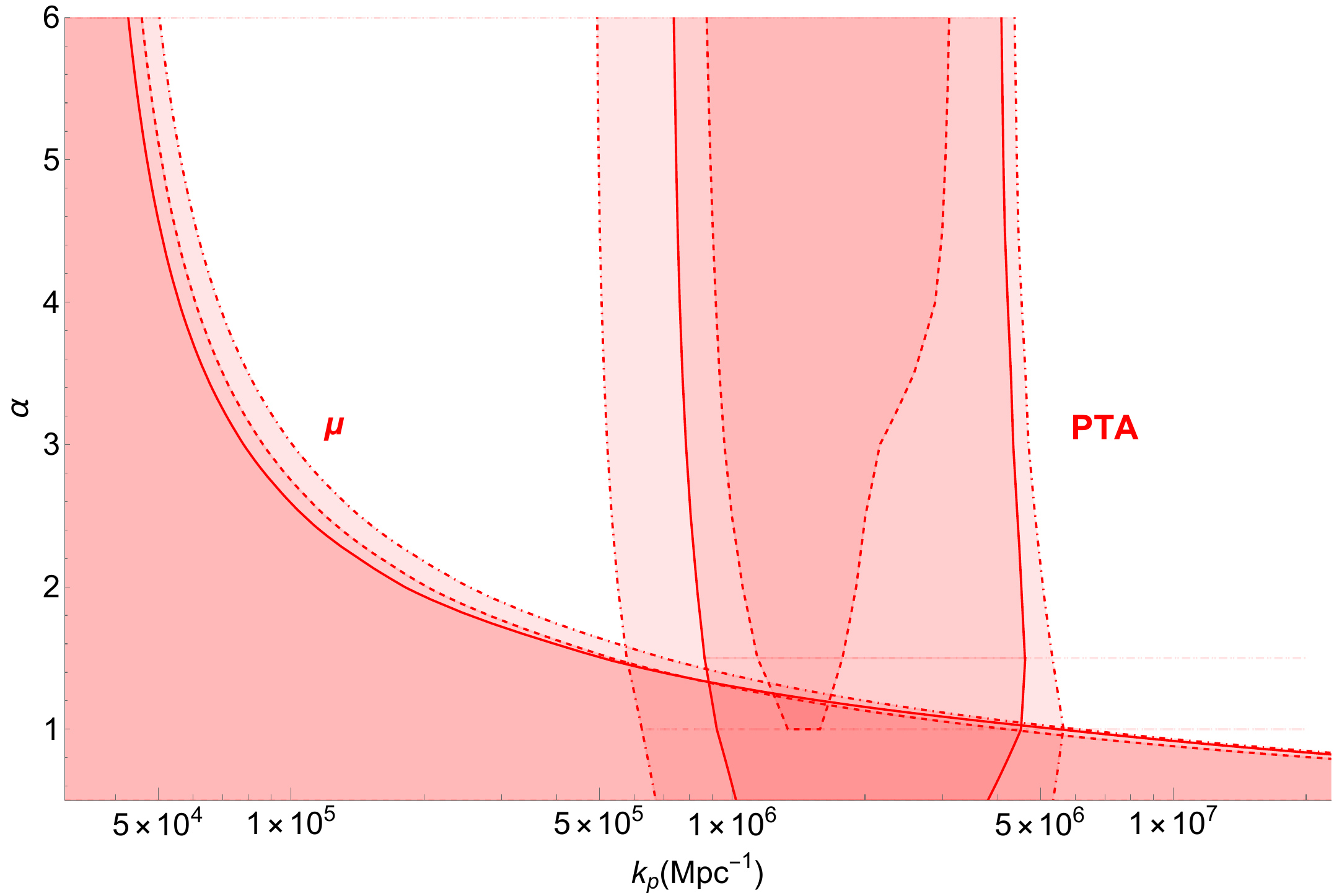}
\end{minipage}
\hfill
\begin{minipage}[b]{0.5\linewidth}
  \centering
  \includegraphics[width=\linewidth]{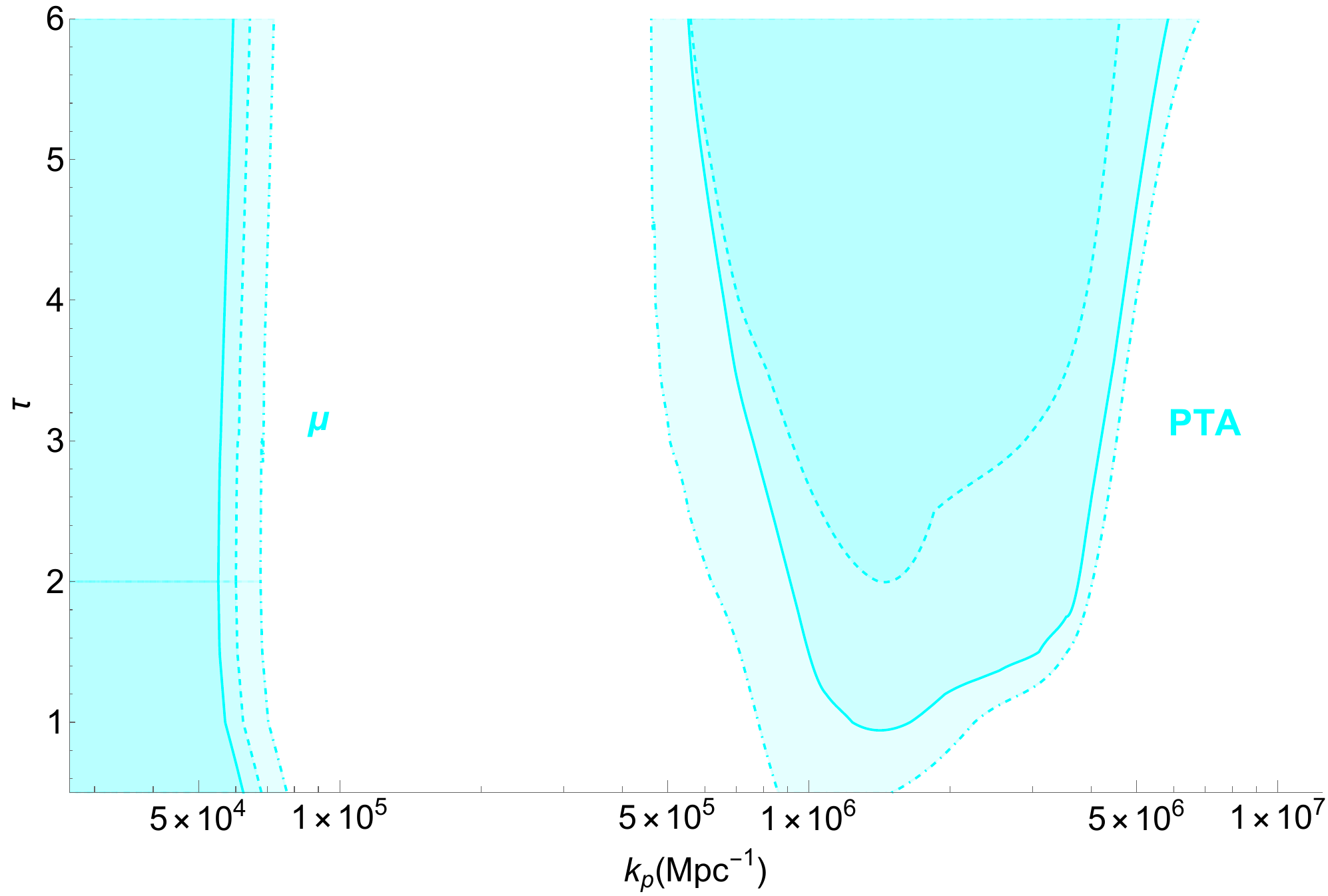}
\end{minipage}
\begin{minipage}{\linewidth}
  \centering
  \includegraphics[width=0.9\linewidth]{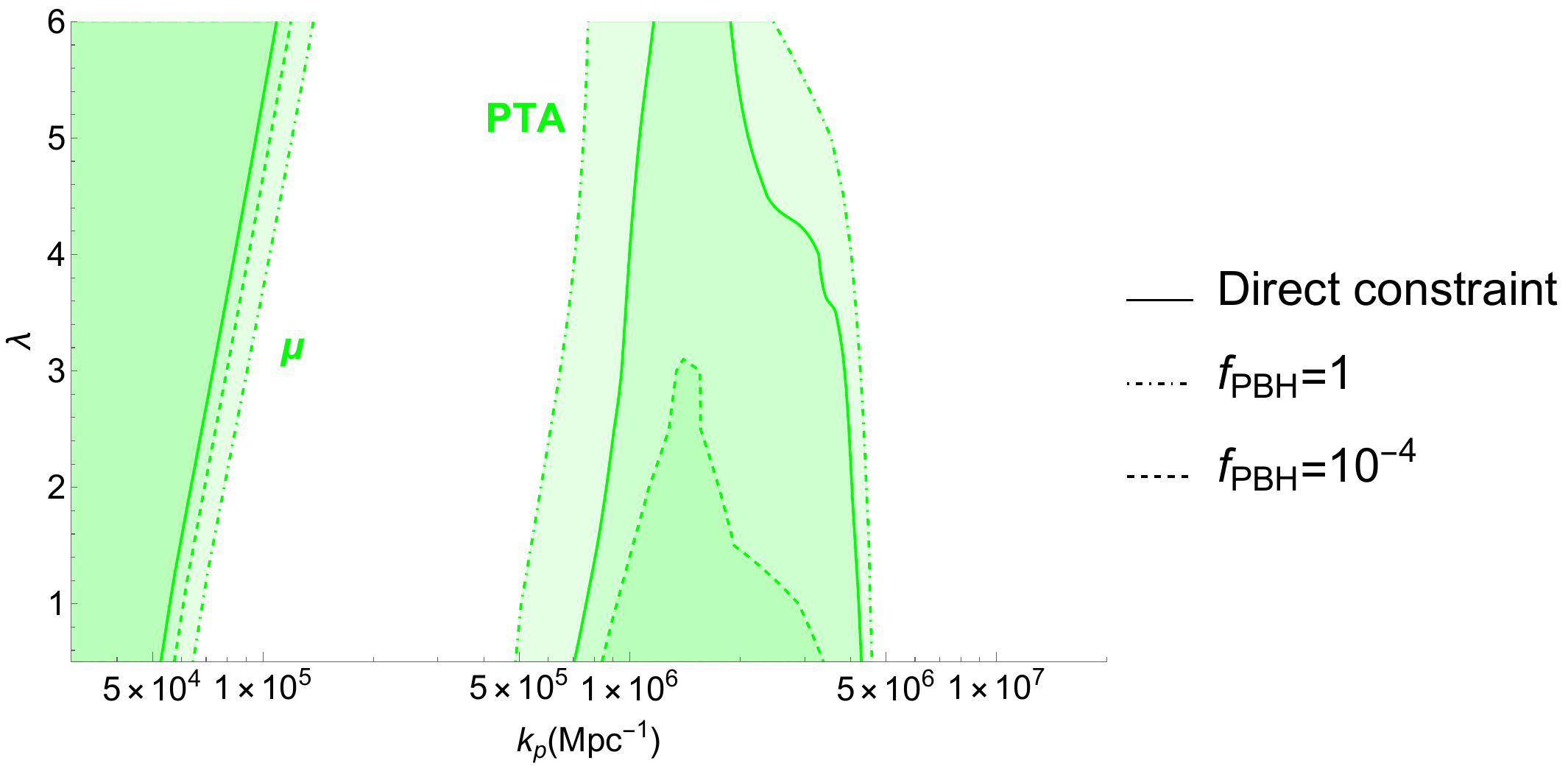}
  \caption{The constraints on BPL parameters ($Top$ $left$: $\alpha$, $Top$ $right$: $\tau$, $Bottom$: $\lambda$) and $k_p$ coming from $\mu$-distortions and PTA, where shaded regions are ruled out. We have also included $f_{\text{PBH}}=1$, 10$^{-4}$ constraints. When varying each shape parameter respectively, we set the remaining two according to $\alpha=4$, $\tau=3$ and $\lambda=1$.}
\label{BPLconstraintplot}
\end{minipage}
\end{figure}

To produce the parameter space constraint plots for the BPL equivalent to \cref{lognormconstraints}, we had to fix two of ($\alpha$, $\tau$, $\lambda$). For example, in the [$k_p$, $\alpha$] plot we fixed both $\tau$ and $\lambda$. The fixed values we chose were based on \cite{PTApbh}, who found $\alpha=4$, $\tau=3$, and $\lambda=1$ to be a relatively conservative parameter combination. In other words, a small change in each respective fixed parameter won't affect our conclusions. The results are shown in \cref{BPLconstraintplot}.

We can see from \cref{BPLconstraintplot}, that for shape parameters $\tau$ and $\gamma$ (describing the decay and width of the BPL power spectrum) the variation in constraining power of $\mu$-distortion observations is not large. We note that this is particularly evident for the $\tau$ parameter, which is to be expected as this describes the tail of the distribution. For the growth parameter, $\alpha$, we see that this is not the case. Decreasing $\alpha$ can be thought of as flattening out the power spectrum on scales $k<k_p$. Thus, it is also expected that decreasing $\alpha$ will lead to tighter $\mu$-distortion constraints on power spectra peaked at smaller scales. This can be seen from the plot, which suggests a difficulty motivating BPL spectra with $\alpha\lesssim1$. 

Similar to the lognormal constraint plot, $\mu$-distortion constraints appear to be insensitive to the value of $f_\text{PBH}$. On the other hand, we find that if we take $f_\text{PBH}=10^{-10}$ (as was done in the lognormal case) then PTA null-detections rule out none of the parameter space. We thus compare constraints with the power spectrum amplitude leading to $f_\text{PBH}=10^{-4}$. Using these constraints we consider the BPL parameter space permitting multiple PT imprints on PBH abundance in \cref{Appendix - BPL}. To make a more direct comparison between the lognormal we focus on the $\lambda$ (width) parameter of the BPL for the remainder of this subsection. 

As mentioned in \cref{variancesection}, increasing the width parameter for both considered power spectra affects the variance of perturbations with respect to the density contrast differently. Direct PBH constraints depend on this quantity (as can be seen from rearranging \cref{betaConstraints}), making this an important distinction between the two spectra. Notably, increasing the width of the BPL (lognormal) results in more (less) tight direct constraints on the power spectrum amplitude. For the case of a BPL power spectrum with $\lambda\gtrsim$ 6, we find that PTA constraints are no longer comparable to direct observational constraints. In other words, direct observations more tightly constrain the power spectrum amplitude than PTA constraints. If we had used log-uniform strain data there would be less unconstrained parameter space. However, we don't expect our conclusions regarding the BPL power spectrum to depend on this choice. 

From \cref{BPLconstraintplot}, we can see the other parameter space constraints don't share this property, with PTA constraints remaining relatively similar upon increasing $\alpha$ and $\tau$. Looking at \cref{BPLratio}, we see that a large portion of \say{interesting}  (as defined in \cref{ratiosection}) parameter space is unconstrained by $\mu$-distortion and PTA constraints. Thus, the shape of the power spectrum plays an important role if we wish to constrain the impact of the QCD PT on the PBH abundance using indirect observations. We have also shown that the constraining power of both direct and indirect observations can differ greatly based on the assumed peaked power spectrum shape. Upon variation of power spectrum parameters, the behaviour of the variance, $\sigma_0^2$, plays an important role in this reasoning. We have shown that these differences in behaviour become more apparent when increasing the width of our respective power spectrum. 

\section{Conclusions}\label{section 5}

In this paper, we have calculated constraints on two popular parametrisations of peaked power spectra. Combining direct observational non-detections of primordial black holes, recent pulsar timing array data, and $\mu$-distortions we have put constraints on the parameters of each power spectrum. We have also compared this to $\mathcal{P}_\zeta$ constraints assuming different values of $f_\text{PBH}$. We find the lognormal power spectrum parameter space to be more tightly constrained than that of the broken power law, offering less flexibility in describing populations of primordial black holes. We believe the additional freedom of the broken power law comes from having extra parameters and sharper cut-offs about the peak. 

We have paid particular interest to standard model phase transitions throughout our analysis, investigating the impact that the corresponding drop in $\omega$ (and thus threshold for collapse) has on $f(M_\text{PBH})$. This was in part due to each phase transition scale corresponding to an interesting cosmological scale such that if a population of primordial black holes were to form around said scale, it could offer a solution to a cosmological unknown. In the context of the lognormal power spectrum, we find it difficult to motivate any of these scenarios.

For the broken power law we find that, upon increasing the width of the power spectrum, the variance of perturbations with respect to the density contrast decreases. This, in turn, leads to tighter direct PBH constraints on the power spectrum amplitude. We note that this also lowers the required amplitude needed to generate a given abundance of primordial black holes. We find that for width parameter $\lambda\gtrsim6$ the pulsar timing array constraints are no longer competitive to the direct PBH constraints on the power spectrum. This results in a large region of parameter space in which the QCD phase transition has a large impact on primordial black hole abundances, whilst evading pulsar timing array constraints. Thus, different parameterisations of peaked power spectra can lead to vastly different conclusions as to whether the shape of our abundance curves can be impacted by the standard model-related drops in the equation of state. 

Nonetheless, we found that for peaked power spectra it was difficult motivating primordial black holes with the drop in equation of state associated to the electroweak phase transition. We have also discussed the difficulties related to power spectra peaked at scales around $e^-e^+$ annihilation. Moreover, capturing the impact of more than one phase transition on primordial black hole abundances is not permitted for any of the parameter space values we considered. 

Lastly, we show in the appendices that our results are insensitive to several extensions we could make to our calculations. Namely, non-linearities, collapse dynamic effects, and constraints including the full equation of state. This gives us more confidence in our conclusions above. 

\begin{center}
    {\bf Acknowledgements} 
\end{center}

The authors thank A. Gow and S. Young for commenting on a draft of this paper and I. Aldecoa Tamayo, A. Aljazaeri, G. Assant, R. Cereskaite, D. Gillies, A.J. Iovino, R. Munoz, K. Saikawa and S. Shirai for helpful conversations. XP is supported by an STFC studentship. CB is supported by STFC grants ST/X001040/1 and ST/X000796/1. 
\appendix

\section{A broad BPL example}\label{Appendix - BPL}

As mentioned, the $e^-e^+$ annihilation scale coincides with tight $\mu$-distortion constraints. Thus, when asking whether we can have a peaked power spectrum that captures features of two early universe phase transitions we are talking about the QCD and EW crossovers. The BPL offers a more interesting answer to this question than the lognormal, which requires an unreasonable width. As can be seen from  \cref{BPLconstraintplot} there is a large portion of unconstrained [$k_p$, $\tau$] parameter space with small $\tau$. We can thus realise a situation in which two PTs significantly impact our abundance calculations: $k^4$ growth in power \cite{byrnes2} (or steeper \cite{zhai,Tasinato,Carrilho,Ragavendra} to a peak at 10$^5$ Mpc$^{-1}$$\lesssim k_p\lesssim k_{QCD}\approx3\times10^6$ Mpc$^{-1}$ with $\tau\ll1$. 

In this scenario, we are able to capture the effects of the QCD PT on PBH abundances, due to the position of the peak. Combined with the steep growth we are also able to evade $\mu$ constraints. Lastly, the shape parameter $\tau\ll1$ results in a long tail in our distribution, allowing us to also capture EW effects on our abundance. This seems interesting until we consider correlations between $\delta$ and $\zeta$. The correlation coefficient, discussed in the appendix of \cite{Young1}, is defined as
\begin{equation}\label{CorrelationCoefficient}
    \gamma_{\delta\zeta}=\frac{\sigma_{\delta\zeta}^2}{\sigma_\delta\sigma_\zeta},
\end{equation}
with
\begin{equation}\label{CCsigmas}
\begin{split}
\sigma_{\delta\zeta}^2  & = \frac{4}{9} \int^\infty_0 \frac{\text{d}k}{k}(kR)^2\Tilde{W}(k,R)\mathcal{P}_\zeta(k) ,\\
\sigma_\delta^2 & = \frac{16}{81} \int^\infty_0 \frac{\text{d}k}{k}(kR)^4\Tilde{W}^2(k,R)\mathcal{P}_\zeta(k)   ,\\
\sigma_\zeta^2 & = \int^\infty_0 \frac{\text{d}k}{k}\mathcal{P}_\zeta(k).
\end{split}
\end{equation}
If $\gamma_{\delta\zeta}$ is close to one, then $\delta$ and $\zeta$ are highly correlated. In other words, a peak in $\delta$ corresponds to a peak in $\zeta$. This validates the high-peak limit assumption, and thus spherical symmetry. On the contrary, if the correlation coefficient is close to zero then perturbation theory based on $\zeta$ doesn't work. In other words, whilst small values of $\delta$ could give $f_\text{PBH}\approx1$, this could result in $\zeta$ greater than unity. This leads to our results being unreliable, so we would ideally be using power spectrum parameters such that $\gamma_{\delta\zeta}\approx1$.

\begin{figure}[!ht]
    \centering
    \includegraphics[width=0.75\linewidth]{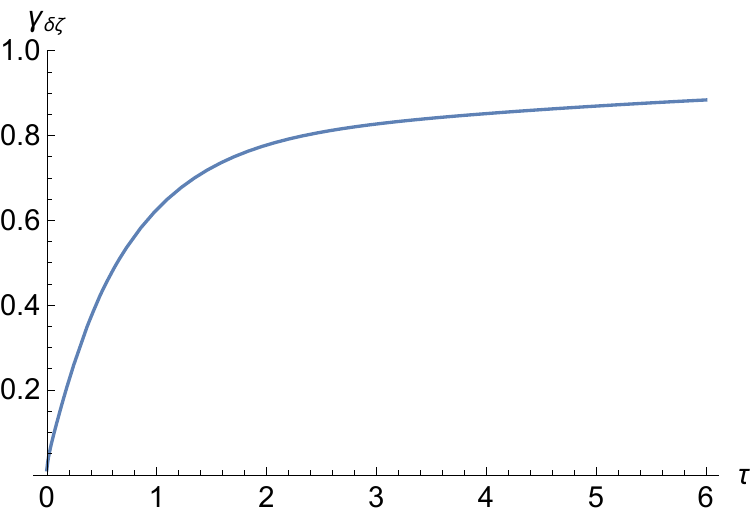}
    \caption{The correlation coefficient, $\gamma_{\delta\zeta}$ for the BPL shape parameter $\tau$. This value quantifies the correlation between $\delta$ and $\zeta$. As $\tau$$\ll1$, these quantities are highly uncorrelated, which would lead our results to be unreliable.}
    \label{correlationcoefficientplot}
\end{figure}

In \cref{correlationcoefficientplot} we plot the correlation coefficient for the $\tau$ parameter of the BPL $\mathcal{P}_\zeta$. The figure shows that for $\tau\ll$1 we obtain $\gamma_{\delta\zeta}\ll1$. Thus, we are unable to draw any reliable conclusions from our method regarding $\tau\ll1$. A log-likelihood analysis of EPTADR2 and NANOGrav15 data, however, has shown that this scenario is largely disfavoured \cite{Franciolini1}.  

We calculate $\gamma_{\delta\zeta}$ for the lognormal and BPL parameter values leading to the $f_\text{QCD}$/$f_{\rm peak}$$\approx9$, as discussed in \cref{ratiosection}. For the BPL, the width parameter $\lambda=40$ leads to $\gamma_{\delta\zeta}\approx0.39$. For the lognormal, the width parameter $\Delta=2.5$ leads to $\gamma_{\delta\zeta}\approx0.3$. This shows that our results leading to this large impact on abundance are more reliable if we assume the shape of the power spectrum is a BPL instead of a lognormal.

\section{Extensions of our work}\label{Appendix-extension}

\subsection{Threshold calculations}

\begin{figure}
    \centering
    \includegraphics[width=0.9\linewidth]{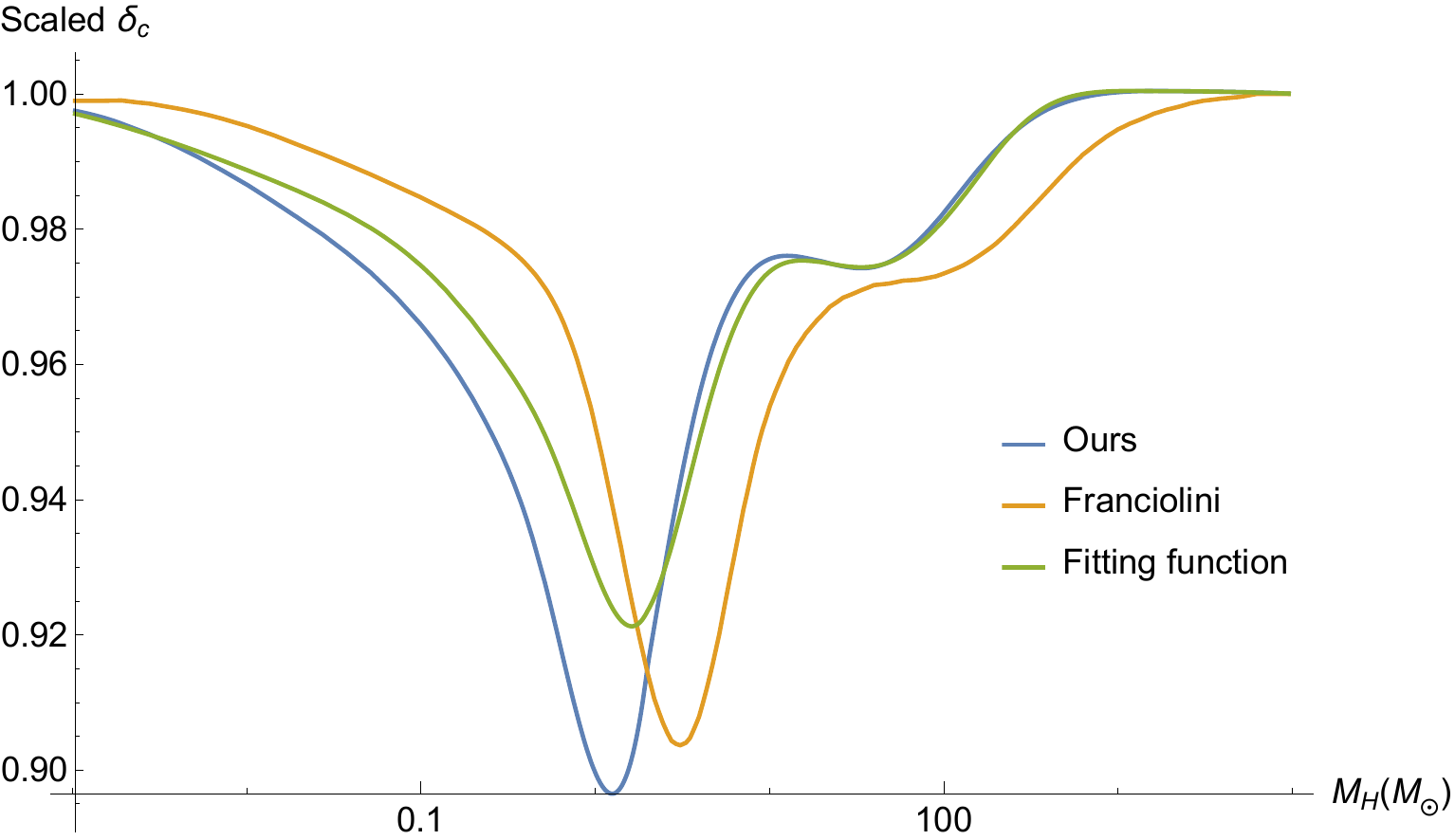}
    \caption{We plot three comparisons of the relative change in threshold for collapse during the QCD PT. We can see that our threshold calculation leads to a similar drop in threshold to \cite{Franciolini}. We also plot results obtained from fitting functions of energy and entropy densities, which leads to a slightly reduced drop.}
    \label{thresholdcomp}
\end{figure}

As mentioned in the main text, when calculating the threshold of collapse for PBHs from the equation of state there are many extensions we can make. Our method, for example, neglects the effects of sound speed and doesn't compute the gravitational potential correctly. Calculating the threshold including these effects was done in \cite{Franciolini}. The shape differs slightly from ours, however the total relative drop is similar. We argue that for our discussions involving the threshold the more important factor is the relative change and not the shape. Thus, we are not concerned by this discrepancy and believe our conclusions remain reliable despite being approximate. We compare these thresholds in \cref{thresholdcomp}.

The third alternative available to us is to rewrite our equation of state in terms of relativistic degrees of energy density, $g_{\ast\rho}$, and entropy density, $g_{\ast s}$. These are related to energy and entropy via the following equations
\begin{equation}\label{gstar}
    g_{\ast\rho}(T)\equiv\frac{30\rho(T)}{\pi^2T^4}, \qquad  g_{\ast s}(T)\equiv\frac{45s(T)}{2\pi^2T^3},
\end{equation}
where $s$ is the entropy density.

We can retrieve the pressure $p$ from the relationship between the energy density and entropy density via $p=sT-\rho$. Then we are able to recast $\omega(T)$ as
\begin{equation}\label{EoS2}
\omega(T)\equiv\frac{p(T)}{\rho(T)}=\frac{4g_{\ast s}(T)}{3g_{\ast\rho}(T)}-1.
\end{equation}
Using fitting functions for these densities, provided by \cite{Saikawa}, we can plot the corresponding threshold for collapse. The lack of reduction in threshold is noticeable, which we expect would lead to different results to ones we found. In particular, to obtain the same impact from a given phase transition on PBH abundance at a fixed $k_p$, we would require a broader power spectrum which is more tightly constrained. 

This is due to the authors specifically constructing smooth functions which reproduce the features of $g_{\ast s}(T)$ and $g_{\ast \rho}(T)$. This does not necessarily mean that fine features in some composite functions of these two quantities can be reproduced. If we wanted to construct composite functions such as $g_{\ast s}(T)/g_{\ast \rho}(T)$ we would require information on the correlations between the two functions, else we have very large errors. We conclude that, in the context of $\omega(T)$ and primordial black hole abundances, it is better to follow the method outlined in the main text.

\subsection{\texorpdfstring{$\kappa(M_H)$, $\gamma(M_H)$}{κ(M\textunderscore{H}), γ(M\textunderscore{H})}}

\begin{figure}[ht!]
    \centering
    \includegraphics[width=0.9\linewidth]{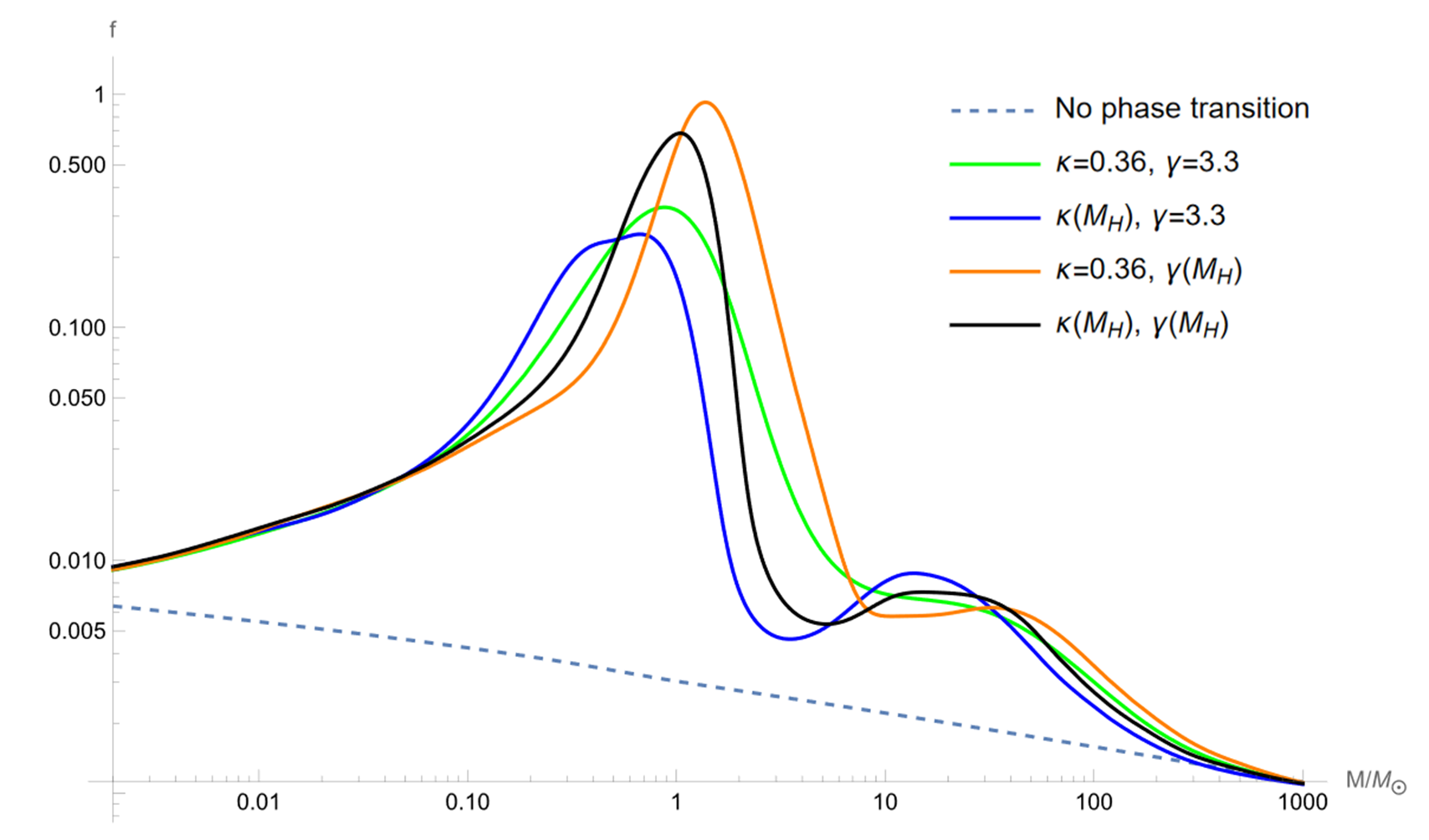}
    \caption{The PBH mass function $f$($M_{\rm PBH}$) calculated varying the $M_H$ dependence of the quantities $\kappa$ and $\gamma$. The dashed line shows the resulting abundance with no PTs. The green line follows our typical calculation.}
    \label{kapgamplot}
\end{figure}

Another extension of our work that could be made is regarding \cref{fPBHMassScaling}. As discussed in \cite{Franciolini}, in a full calculation the quantities $\kappa$ and $\gamma$ are nonconstant during a PT. These numerical factors depend on the radial profile of the perturbations being considered as well as the characteristic scale of the horizon crossing of said perturbations. We plot the impact of the $M_H$ dependence of these quantities on the PBH abundance in \cref{kapgamplot}, assuming a nearly scale-invariant curvature power spectrum.

As can be seen from the plot, the shape of each abundance curve remains similar. There are, however, some interesting distinctions. In particular, accounting for variation in $\gamma$ boosts PBH abundance by a factor of 2. Secondly, varying $\kappa$ seems to shift the peak to slightly lower masses. Once more, we believe our conclusions would not be altered by the inclusion of the variation of these parameters in our calculations.

\subsection{Constraint dependence on $M_H$}\label{constraintMH}

\begin{figure}
\begin{minipage}[b]{\linewidth}
  \centering
  \includegraphics[width=0.8\linewidth]{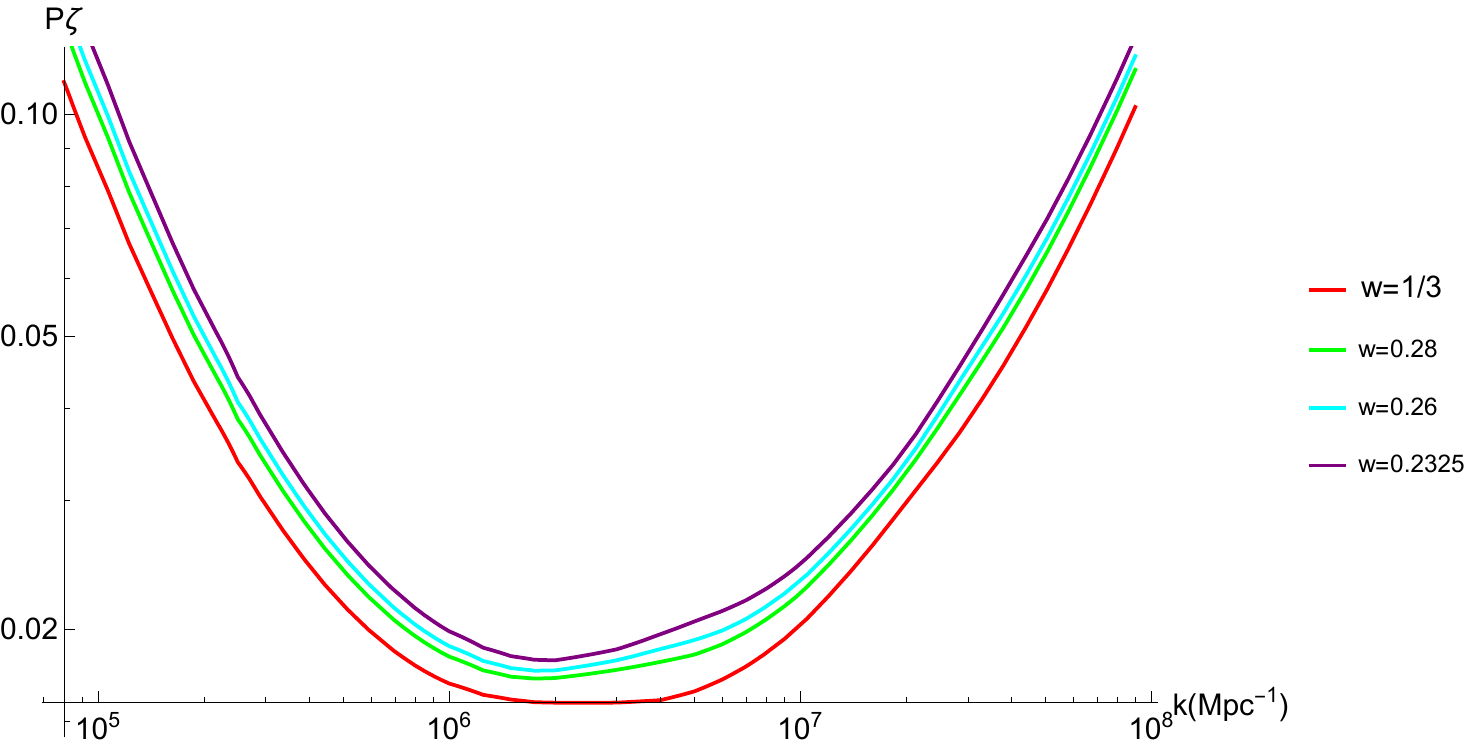}
\end{minipage}
\hspace{3cm}
\begin{minipage}{\linewidth}
  \centering
  \includegraphics[width=0.8\linewidth]{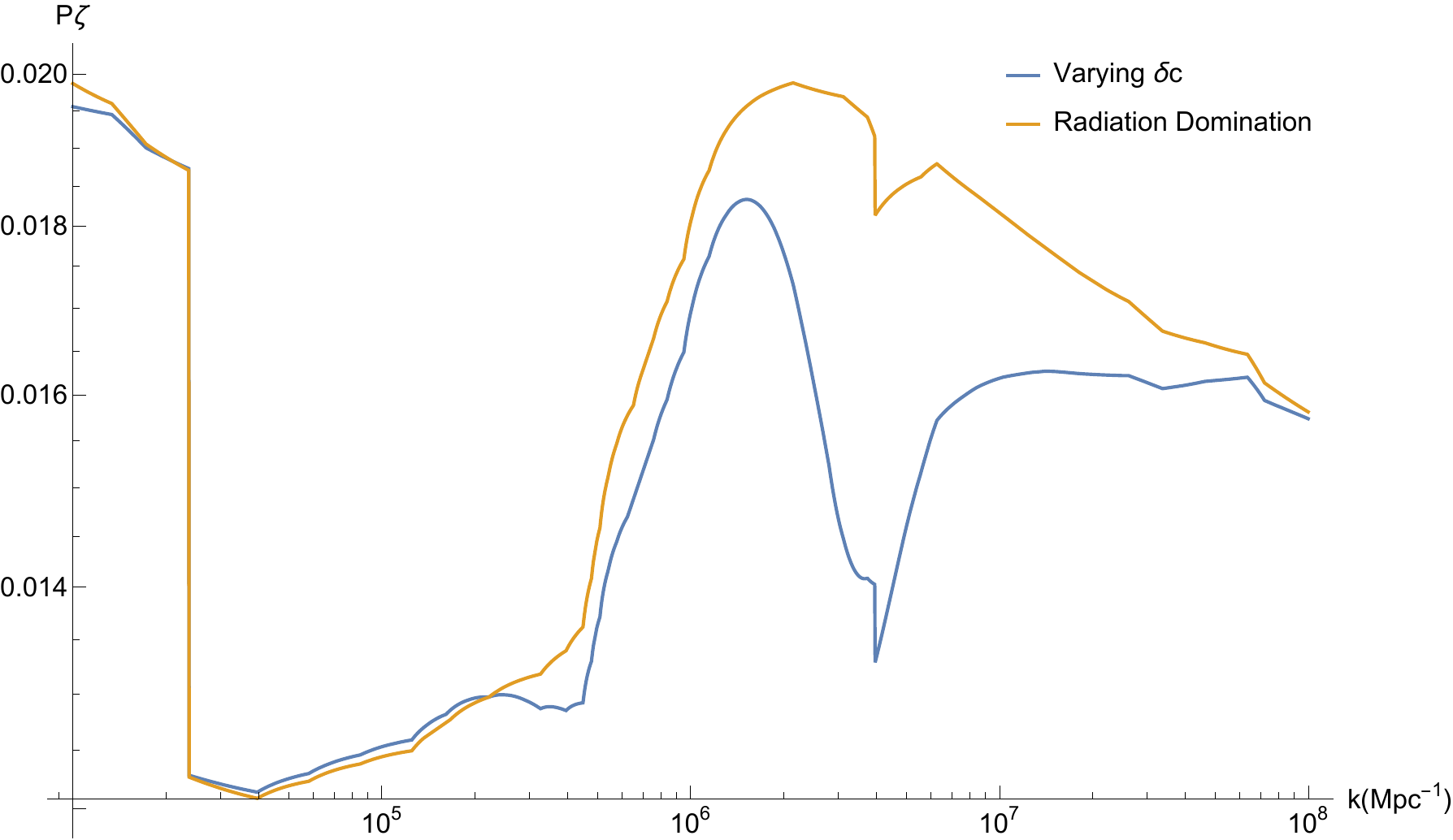}
  \caption{This plot shows the ($Top$: PTA, $Bottom$: PBH) constraints on the lognormal power spectrum, varying the equation of state parameter. As we can see, a reduction in the equation of state results in a weaker constraint on the power spectrum from PTA data, but tighter constraints from the null-detection of PBH. In both cases, the change is relatively small.}
\label{varydcconstraints}
\end{minipage}
\end{figure}

Lastly, we comment on the impact of a varying equation of state on our constraints. For example, the scalar-induced GW spectrum we calculated assumes the constant radiation domination value of $\omega$. As previously discussed, this is not the case in the early universe. We show the PTA constraints for different constant values of $\omega$ in \cref{varydcconstraints}. As can be seen, if $\omega$ decreases the constraints on $\mathcal{P}_\zeta$ are less severe. We can go further toward the complete picture by using $\omega$($T$), however, this makes the computation much more involved. From the figure we expect these changes to be relatively small. We note that this has been done for a cut power-law power spectrum \cite{Abe}.\\

As well as PTA constraints depending on the equation of state, direct PBH constraints also depend on this quantity. This can be seen from \cref{PressSchectbeta}, where it is understood that the threshold depends on $\omega$. Moreover, the overall factor we multiply our PBH constraint line in \cref{Appendix-NL} in order to approximate non-linearities depends on both threshold and gravitational potential. In \cref{varydcconstraints}, we also show the impact of including these additional contributions to the PBH constraint line. Again, we see the impact of this inclusion is relatively small.

\section{Non-linearities}\label{Appendix-NL}

It has been known for some time that the density contrast and curvature perturbation have a non-linear relationship \cite{Harada}. In the super-horizon regime, this can be expressed as \cite{Young:2024jsu}
\begin{equation}
    \delta(\hat{r})=-\frac{4(1+\omega)}{5+3\omega}\left(\frac{1}{aH}\right)^2\text{exp}\left(-\frac{5\zeta(r)}{2}\right)\nabla^2\text{exp}\left(\frac{\zeta(r)}{2}\right),
\end{equation}
where $\hat{r}$ is the radial coordinate in the comoving synchronous gauge, and in spherical symmetry
\begin{equation}
    \nabla^2=\frac{\partial^2}{\partial r^2}+\frac{2}{r}\frac{\partial}{\partial r}.
\end{equation}
The result of this relationship is that we expect the density contrast to be non-Gaussian, regardless of whether $\zeta$ follows a Gaussian distribution. However, when calculating PBH abundances we are more interested in the smoothed density contrast. This can be written in terms of a linear, Gaussian component $\delta_l$ as follows
\begin{equation}
    \delta_m=\delta_l-\frac{3}{8}\delta_l^2.
\end{equation}
The critical amplitude of this component can be calculated by inverting the above, giving
\begin{equation}
    \delta_{c,l\pm} = \frac{4}{3} \left( 1 \pm \sqrt{\frac{2-3\delta_c}{2}}\right).\label{dclminus}
\end{equation}
We can further note that the number density of sufficiently rare peaks in a comoving volume for a random Gaussian field is \cite{Bardeen}
\begin{equation}
 \mathcal{N}=\left(\frac{\sigma_1^3}{4\pi^2\sigma_0^3}\right)\nu^3\text{exp}\left(-\frac{\nu^2}{2}\right),
\end{equation}
where $\nu$$\equiv\delta_l/\sigma_0$. Following \cite{Young2}, peaks theory is then used to calculate $\beta$ such that the non-linear relationship between $\delta$ and $\zeta$ is accounted for, given as
\begin{equation}
    \beta(M_H)=\int^{\frac{4}{3\sigma_0}}_{\nu_{c-}}\text{d}\nu\frac{\kappa}{3\pi}(\nu\sigma_0 -\frac{3}{8}(\nu\sigma_0)^2-\delta_c)^\gamma\left(\frac{\sigma_1}{aH\sigma_0}\right)^3\nu^3\text{exp}\left(-\frac{\nu^2}{2}\right),
\end{equation}
where $\nu_{c-}\equiv\delta_{c,l-}$/$\sigma_0$.

To account for non-linearities when calculating our PBH constraints, we follow the arguments of \cite{Gow}. They assume that each peak that collapses to form a PBH has an amplitude close to the critical value. Using the Fourier transform of the top-hat window function, which provides an analytic relation between $\delta$ and $\zeta$, $\delta_c=0.55$ \cite{Young3} which leads to the critical amplitude of the Gaussian component $\delta_{c,l-}\approx0.77$. Thus, they conclude that peaks should have an amplitude 1.41 times larger if we are to approximate non-linearities. The overall factor we times our PBH constraint line is then 1.41$^2$=1.98, because $\mathcal{P}_\zeta \propto \delta_c^2$.

\printbibliography

\end{document}